\documentclass[sigconf,letterpaper]{acmart}

\setcopyright{none}
\renewcommand\footnotetextcopyrightpermission[1]{}

\usepackage[utf8]{inputenc}
\usepackage{hyperref}
\usepackage{balance}
\usepackage{booktabs}
\usepackage[labelformat=simple]{subcaption}

\usepackage[inline]{enumitem}
\usepackage{standalone}
\usepackage{tikz}
\usetikzlibrary{calc,colorbrewer,decorations,decorations.pathreplacing,fit}

\usepackage{fontawesome}

\usepackage{algorithmic}
\usepackage{graphicx}
\usepackage{textcomp}
\usepackage{xcolor}
\usepackage{float}
\usepackage{caption}
\usepackage{subcaption}
\def\BibTeX{{\rm B\kern-.05em{\sc i\kern-.025em b}\kern-.08em
    T\kern-.1667em\lower.7ex\hbox{E}\kern-.125emX}}

\newcommand{\todo}[1]{\textcolor{red}{\textbf{TODO:} #1}}

\usepackage{pifont}

\usepackage{xspace}
\newcommand{\etal}{\textit{et al.}~}
\newcommand{\eg}{\textit{e.g.,}~}
\newcommand{\ie}{\textit{i.e.,}~}

\newcommand{\one}{({\em i})\xspace}
\newcommand{\two}{({\em ii})\xspace}

\newcommand{\ipbleconn}{\textit{6BLEMesh}\xspace}
\newcommand{\ipbleadv}{\textit{IP-BLE-Adv}\xspace}

\makeatletter
\renewcommand{\paragraph}[1]{\vspace*{0.03in}\noindent{\bf #1.}\hspace{0.25ex \@plus1ex \@minus.2ex}}
\makeatother

\begin{document}

\date{}

\title{IPv6 over Bluetooth Advertisements:\\An alternative approach to IP over BLE}

\author{Hauke Petersen}
\email{hauke.petersen@fu-berlin.de}
\affiliation{
  \institution{Freie Universit\"at Berlin}
  \country{Germany}
}
\author{János Brodbeck}
\email{brodbeck@zedat.fu-berlin.de}
\affiliation{
  \institution{Freie Universit\"at Berlin}
  \country{Germany}
}
\author{Thomas C. Schmidt}
\email{t.schmidt@haw-hamburg.de}
\affiliation{ 
  \institution{HAW Hamburg}
  \country{Germany}
}
\author{Matthias W{\"a}hlisch}
\email{m.waehlisch@fu-berlin.de}
\affiliation{ 
  \institution{Freie Universit\"at Berlin}
  \country{Germany}
}


\begin{abstract}
The IPv6 over Bluetooth Low Energy (BLE) standard defines the transfer of IP data via BLE connections.
This connection-oriented approach provides high reliability but increases packet delays and requires substantial overhead to manage BLE connections.
To overcome these drawbacks we present the design and implementation of IPv6 over BLE advertisements, a standard-compliant connection-less approach.
We deploy our proposal on low-power IoT hardware and comparatively measure key network performance metrics in a public testbed.
Our results show that IP over BLE advertisements offers network performance characteristics complementary to IP over connection-based BLE, trading lower reliability for shorter~latency.
\end{abstract}

\maketitle
\pagestyle{plain}


\section{Introduction}
\label{sec:introductiion}
The Internet of Things (IoT) is highly fragmented~\cite{akgwy-ftsit-18}.
In the low-power wireless IoT, heterogeneous link layer technologies compete, each requiring dedicated (smart) gateways to connect to the Internet.
Bluetooth Low Energy (BLE) is the most deployed low-power radio technology today \cite{btmu-bcsv-20} and the IP over BLE standard~\cite{RFC-7668,bipsp-bcsv-14,draft-gomez-6lo-blemesh-10} allows to seamlessly connect BLE devices to the Internet.
Furthermore, BLE offers best in class low-power characteristics \cite{sarb-llkb-16, ee6ib-dg-20, befpi-sbzr-17} as well as reliable network performance \cite{psw-mgmio-21}, making it a promising default link layer in the low-power IoT.

IP over BLE, however, works on top of BLE connections, which leads to some disadvantages.
First, before exchanging IP data, any node must open BLE connections to one or more adjacent peers.
Managing these connections automatically poses overhead on BLE nodes.
Second, the number of concurrent BLE connections is typically limited to $\approx$15~peers due to restrictions in memory and radio scheduling.
Third, the current IP over BLE standard increases packet delays as BLE connections are time-slotted.
Typical latencies of IP over BLE networks are, in some scenarios, substantially larger compared to networks based on carrier-sense multiple access (\eg IEEE802.15.4)~\cite{psw-mgmio-21}.

In order to mitigate the disadvantages of the current IP over BLE standard, we propose to explore the transfer of IP data using the connection-less mode of BLE.
We do not aim for replacing the existing connection-oriented IP over BLE design but to offer an alternative based on the same technology, to allow IoT developers to optimize deployments depending on requirements.
As both designs are based on the same software (BLE and IP stacks) and hardware (radios) they can be deployed and run simultaneously, which finally will increase IoT use cases for BLE networks.

In this work, we present the protocol design and prototype implementation of IPv6 data over connection-less BLE.
We utilize the extended advertisements, which were introduced in Bluetooth version~5.0~\cite{b50-bcsv-16}.
Extended advertisements have the advantage of offering a MTU of up to 65~Kbytes through packet fragmentation capabilities provided by BLE controller.
They are, thus, able to carry full IPv6~packets with a minimum MTU of 1280~bytes on top of a lean software system.
In contrast to this, legacy BLE advertisements would allow only for a maximum payload of 31~bytes per packet, which would require complex fragmentation schemes implemented on an intermediate layer between IP and BLE in addition to 6LoWPAN-based header compression~\cite{RFC-4944,RFC-7400}.

We systematically measure key performance metrics in practice based on 15 low-power BLE nodes in the FIT IoTlab testbed~\cite{abfhm-filso-15}.
We analyze reliability, latency, and energy consumption in different single- and multi-hop network topologies and compare them to the performance of connection-based IP over BLE networks.
Our results show that using advertisements offers lower latency (on average 1.5$\times$ to 5$\times$ lower for comparable configurations) but less reliability (1\% to 80\% packet loss vs $<$0.01\%) and increased power consumption (radio always on).

Currently, Bluetooth Mesh \cite{bmp-bcsv-19} is the only standard to transfer (proprietary) data over connection-less BLE.
In contrast to our proposal, however, Bluetooth Mesh does not support arbitrary IP~packets but is limited to the flooding of specific, small data frames (<20~bytes), and does not support fragmentation.
Bluetooth Mesh aims for vendor-specific simplified scenarios, whereas our proposal targets flexible Internet-like deployments.

In summary, we make the following contributions:
\begin{enumerate}[itemsep=0pt]
\item The first, standard-compliant design to transfer IPv6 data over BLE extended advertisements. (\autoref{sec:protocol-design})
\item A publicly available, open source implementation based on the operating system RIOT and the BLE stack NimBLE. (\autoref{sec:system_design})
\item Reproducible experiments conducted on real-world hardware, and all artifacts. (\autoref{sec:eval})
\item A comparative performance evaluation including network and system measures to show protocol mechanics in contrast to connection-based IP over BLE networks. (\autoref{sec:results}--\autoref{sec:discussion})
\end{enumerate}



\section{Background}
\label{sec:background}

BLE supports three modes to transfer data: the connection-less \emph{legacy advertising mode}, the connection-less \emph{extended advertising mode}, and the \emph{connection-based mode}.
Connection-less communication is usually used to enable the discovery of services and to broadcast data for further processing to unknown peers.
Connection-based communication aims for communication between direct peers, \eg in the IP over BLE standard \cite{bipsp-bcsv-14,RFC-7668}.
This section briefly presents core background on all three modes with a focus on embedding data.

\subsection{Connection-less BLE Communication}

Legacy advertising is used in Bluetooth Mesh \cite{bmp-bcsv-19} and the extended advertising mode was introduced in Bluetooth 5.0 \cite{b50-bcsv-16}.

\begin{figure}
  \centering
  \includegraphics[width=.45\textwidth]{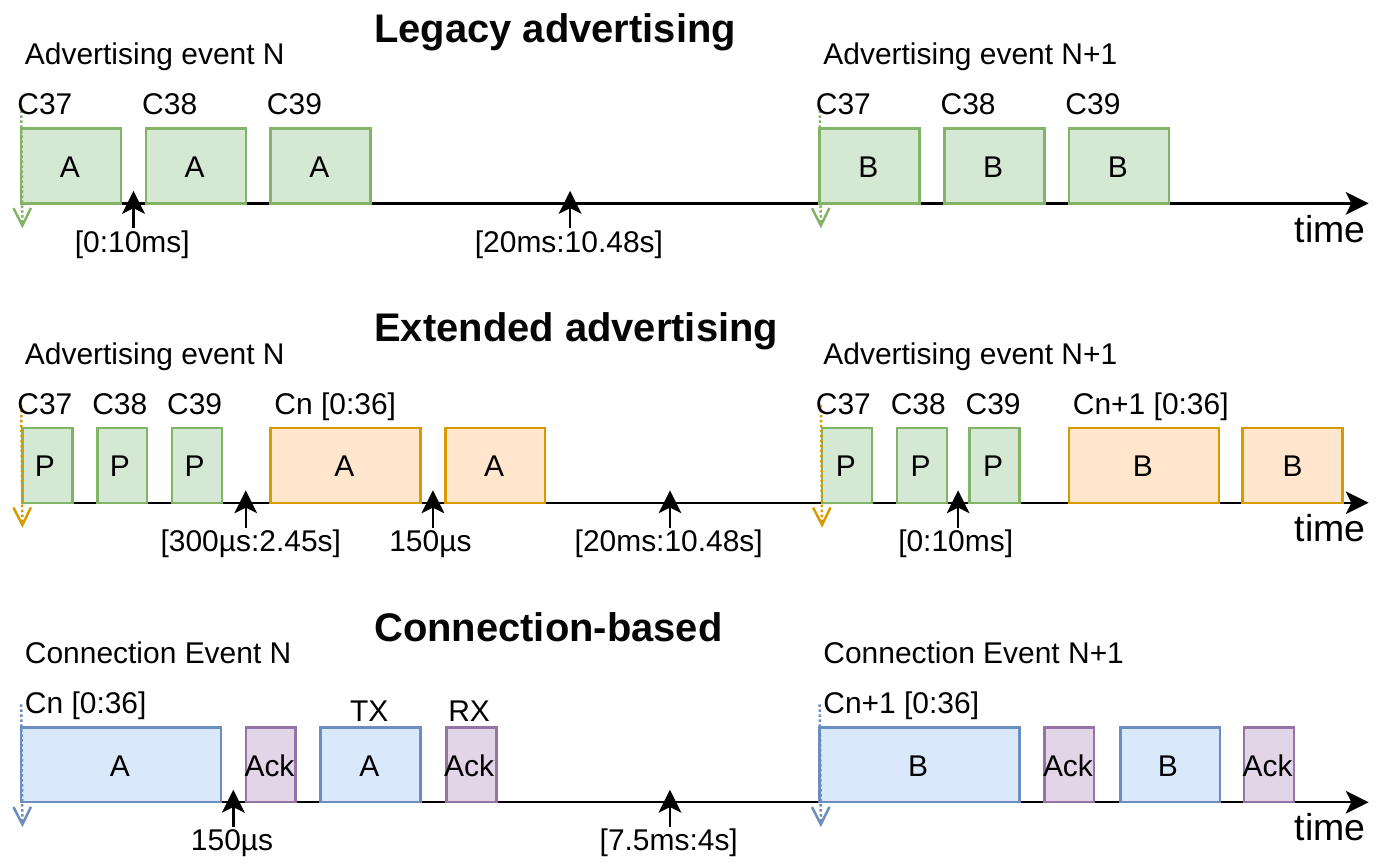}
  \caption{BLE packet flow when transferring payloads A and B using legacy and extended advertisements as well as connection-based communication.}
  \label{fig:pktflow}
\end{figure}

\paragraph{Legacy Advertising}
Legacy advertising supports a maximum payload of 31~bytes.
The advertisement packets are sent periodically in so-called \emph{advertising events} during an \emph{advertising interval}, depending on the configuration between 20ms and 10.48s, see \autoref{fig:pktflow}.
Each advertisement packet is sent via the \emph{primary advertisement channels}, three predefined channels that are exclusively reserved for advertisements to achieve some level of robustness.
This mode is unidirectional (no link layer acknowledgements) and unmanaged (no CSMA scheme).


To receive advertisements, nodes listen periodically on one of the primary advertising channels (\textit{scan event}) based on the \textit{scan interval}.
An active radio in RX mode is expensive in terms of energy.
The Bluetooth standard, thus, allows the receiver during each scan event to only activate the radio during the \textit{scan window}.
If the scan window is shorter than the advertising interval, advertising packets might get lost.
There exist a number of approaches on how these parameters can be optimized to balance energy usage and delivery probability
\cite{sedds-sh-21, pmand-sr-20, bpcnd-jlmp-17}.

In common advertising use-cases (\eg beaconing), devices use a fixed payload.
When considering less predictable application data such as carried in IP packets, this data should preferably be sent within a single advertising event.
Since there is no guarantee that an advertising event is received, a single payload is typically transmitted in multiple advertising events, hence implementing a fixed number of link layer packet retransmissions.
This is, for example, applied in Bluetooth Mesh, which defaults to carry the same payload in 5~connection events \cite{bmp-bcsv-19}.

When transmitting IP data using legacy advertisements, the limited payload becomes a major bottleneck.
To encapsulate IP~data into the payload of advertising packets, an additional advertising data header of at least 6~bytes is required, leaving only 25~bytes for IP data.
Even when using header compression techniques (\eg defined in 6LoWPAN~\cite{RFC-6282, RFC-7400}), packet fragmentation would then be needed.

\paragraph{Extended Advertising}
Extended advertising allows for larger payloads and is based on legacy advertising.
Instead of carrying data in (very limited) packets via the primary channels, extended advertising uses these packets to refer to one of 37~data channels and a start time.
The actual payload is then sent at the specified start time on the given data channel in one or more chained data packets (see \autoref{fig:pktflow}).

By containing only a short pointer, the packets sent on the 3 advertising channels need less air time for transmission.
This reduces collision probability on those potentially crowded channels.
The collision probability for data packets is reduced by utilizing all 37~data channels for their transmission.

By splitting the payload over multiple chained packets, extended advertising allows to transfer up to 65~Kbytes in a single advertising event.
Fragmentation and reassembly into link layer data packets is done by the Bluetooth controller, which relieves higher layers from implementing fragmentation schemes to transfer full IPv6 MTUs.

\subsection{Connection-based BLE Communication}
In contrast to advertisements, which are transmitted in the broadcast domain, BLE connections are always point-to-point.
In a connection between two nodes, one node acts as \textit{connection coordinator} while the peer node is the \textit{connection subordinate}.\footnote{The terms ``coordinator'' and ``subordinate'' used in this paper diverge from Bluetooth specifications, to support non-discriminatory language.}
Similar to advertising events, the communication in the connection-based mode is structured into \textit{connection events}.
Each connection event consists of at least a single data packet exchange between the coordinator and subordinate.
This can be repeated multiple times in the same event until no payload is left to send or the next connection event starts.
If one of the peers has no data to send, it will send empty packets.
Each connection event takes place on one of the 37 available data channels. 

BLE connections provide a point-to-point service guaranteeing first-in-first-out, in-order, and complete data delivery.
To achieve this, data packets are retransmitted on the link layer until they are acknowledged.
If by either side no valid packet is received during a specific amount of time, the connection is considered lost and is closed.
Consequently, as long as connections are active, there is no packet loss on the link layer~\cite{psw-mgmio-21}.
The IP over BLE standard~\cite{bipsp-bcsv-14,RFC-7668,draft-ietf-6lo-blemesh} is using this connection-based mode.

\section{IPv6 over BLE Extended Advertisements}
\label{sec:system}
This section describes our design to enable connection-less IPv6~communication over BLE.
The core idea is to carry IPv6 packets in the payload of BLE extended advertisements, using either directed advertisements or undirected advertisements to transmit unicast or multicast data, respectively.

In the remainder of this paper, we will denote our proposed design \ipbleadv, while the standardized IP over connection-based BLE is denoted \ipbleconn.




\subsection{Protocol Design}
\label{sec:protocol-design}

\paragraph{Requirements}
Wherever applicable, our proposal shall comply with the \ipbleconn standard.
In detail,
\begin{enumerate}
  \item the support of an MTU of $\ge$ 1280~bytes across all links to prevent the fragmentation of IPv6~packets~\cite{RFC-8200} and to utilize the build-in functionality of extended advertisements.

  \item the use of 6LoWPAN~header compression~\cite{RFC-4944}.

  \item the support of unicast and multicast messages, where multicast messages can be transmitted via broadcast (similar to IEEE~802.15.4) but unicast messages should be filtered on the link layer.
\end{enumerate}


\paragraph{Advertising Types} All data is transferred using BLE extended advertisements.
BLE supports different types of extended advertisements. 
In our design, all advertising packets are non-connectable and non-scannable.
To maximize the use of build-in functionality of the BLE controller, we suggest to use non-directed packets to transfer multicast messages and directed advertising packets for unicast messages.
The latter allows to utilize build-in packet filtering of an BLE controller, which prevents the need to filter on the upper layer and thus saves processing and energy.

\paragraph{Advertising Data Encoding} The Bluetooth standard requires the payload of advertising packets to be encoded in the advertising data~(AD) format.
This format comprises a list of one or more generic length-type-value fields with a 1~octet length field, a 1~octet type field, and a variable length data field.
The structure of the data field depends on the value of the type field \cite{bcss-bcsv-21}.

Encoding IP data packets into this format poses two challenges: \one there is no predefined AD type to carry IP data and \two the maximum length of a single AD field is limited to 254 bytes.
To address \one, we opted to use the AD type \texttt{Service Data - 16 bit UUID}.
In this type, the data section of the AD field starts with a 2~octet Bluetooth service UUID and is followed by the IP~payload including the prepended sequence number.
Due to the lack of a standardized service identifier for \ipbleadv we use a custom 16-bit UUID.

This design leaves 252~bytes of IP data in each AD segment.
If an IP packet exceeds this size, it is split into multiple chained AD~segments of the same type, where each segment is filled with the maximum possible amount of data before a new segment is started.
As a result, each IP packet to be transferred, including a custom sequence number (see below), is encoded into a continuous list of chained AD~segments, which can be passed in a single block to a BLE stack.
Fragmentation and reassembly of this block is implemented transparently by the BLE stack.

\paragraph{Data Reception} All nodes are expected to constantly listen for incoming packets to maximize reception reliability.
This behavior is also specified for Relay nodes in Bluetooth Mesh or for nodes in CSMA/CA-based IEEE802.15.4 modes.
It implies that the scan interval and scan window are equal.

Having nodes in a always-on radio state per default conflicts with energy requirements, but it is sufficient to gain insights on the network performance of \ipbleadv networks.
Looking into concepts to improve energy efficiency by duty cycling the radio, like the friend role defined by Bluetooth Mesh, would be desirable.
This is however not in scope of this work.

\paragraph{Data Transmission} 
A single extended advertising event consists of 3~advertising packets sent on each of the three advertising channels, as well as 1 or more chained packets containing the actual payload sent on one of the 37 data channels (see \autoref{sec:background}).
Ideally, each IP packet is sent in a single advertising event.
Due to packet losses on the advertising channels, radio switching delays on the receiver, and the unidirectional nature of advertising packets, there is no guarantee that peer nodes receive a packet.
Repeating the same payload in multiple consecutive advertising events does significantly improve the packet delivery ratio.
Our design offers to configure this number of static retransmissions in the same style as Bluetooth Mesh does this for Relay nodes.
In \autoref{sec:results_params} we present our findings towards optimizing this parameter.

\paragraph{Duplicate Detection} By retransmitting IP packets through multiple connection events, nodes potentially receive the same IP data packet multiple times.
As we do not require duplicate detection on upper layers, BLE~stacks must be able to detect these duplicates.
The Bluetooth standard defines means for duplicate detection on advertising packets in the BLE controller.
In practice, this duplicate detection is implemented using buffer memory holding the latest received packets to be compared with newly received ones.
Especially on memory constrained devices, this buffer space is limited.
We noticed that in environments with a high amount of advertising traffic, these buffers are not able to hold a sufficient amount of data and hence will fail to detect duplicates reliably.

As the build-in duplicate detection is unreliable, we introduce a 1~octet sequence number that is prepended before the IP~payload.
This sequence number is incremented for each IP payload that is transmitted.
Each receiver maintains a table of link layer source addresses and last received sequence number for each neighbor.
Incoming packets are then filtered by comparing the included sequence number against the last seen sequence number of packets from the same source address.
If the sequence numbers are equal, the incoming packet must be a duplicate and is dropped.

\subsection{System Design and Implementation}
\label{sec:system_design}

\paragraph{High-level Idea} Our proposed solution consists of a 6LoWPAN-enabled IP stack connected to a BLE stack, including a wrapper module between the stacks, taking care of forwarding and converting IP in both directions.
In the context of the IP stack, the wrapper module acts as a plain network interface exposing the 6-byte BLE addresses as link layer addresses following the specification in the IP over BLE standard~\cite {RFC-7668}.
The interaction between the wrapper module and the BLE stack is restricted to the \textit{Generic Access Profile}~(GAP).
Next to the BLE controller, on the host side, our design requires only the \textit{Logical Link Control and Adaption Layer Protocol}~(L2CAP) layer to be implemented.
In \ipbleconn, the presence of a defined \textit{Generic Attribute Profile} (GATT) service is required by the \textit{Internet Support Profile} (IPSP) \cite{bipsp-bcsv-14}.
The service aids the establishment of BLE connections, and hence is not needed in our implementation.

\paragraph{Implementation}
We implemented the proposed design in a fully open-source platform based on the RIOT operation system \cite{bghkl-rosos-18,riot-web} and the Apache NimBLE BLE stack \cite{nimble-web}.
For IPv6 networking, we utilize GNRC, the default IPv6 stack in RIOT~\cite{lkhpg-cwemr-18}.

All IP data forwarding as well as link layer address handling is implemented in a single software module called \texttt{jelling}.
In the context of the GNRC network stack, this module acts as a network interface by implementing GNRC's \texttt{netif} interface.
To interact with the NimBLE BLE stack, the proposed module uses NimBLE's GAP API, in particular the \texttt{ble\_gap\_ext} family of functions.

On data reception, the NimBLE GAP API does, depending on the overall packet size, potentially fragment incoming data into chunks and pass those chunks step-by-step to the API user.
The received IP data is encoded into one or more BLE advertising data format segments (see \autoref{sec:system_design}).
However, in order to be able to extract the IP packet, access to the full packet is needed.
To achieve this, our implementation introduces an intermediate receive buffer that holds at least one full IPv6 MTU plus the overhead generated by the advertising data field headers, leading to an additional RAM usage of 1.3~Kbytes.

For \ipbleconn, we use the \texttt{nimble\_netif} implementation that is included in RIOT~\cite{psw-mgmio-21}.


\section{Evaluation Setup}
\label{sec:eval}
We evaluate our proposal (see \autoref{sec:system}) based on low-power hardware.

\paragraph{Hardware} We use the \texttt{nrf52840dk} and \texttt{nrf52dk} development boards from Nordic Semiconductor.
Both feature 64 MHz ARM Cortex-M4F SoCs with on-chip BLE 5.2 support.
In the context of this evaluation, the only relevant differences between both platforms are the RAM and ROM sizes: the \texttt{nrf52840dk} offers 1~Mb ROM and 256~Kb RAM while the \texttt{nrf52dk} offers
512~Kb ROM and 64~Kb RAM.
Both SoCs, especially the \texttt{nrf52dk}, offer common memory and performance characteristics of modern IoT platforms~\cite{RFC-7228}.

\paragraph{Testbed} We perform all experiments at the Saclay site of the FIT~IoTlab \cite{abfhm-filso-15} using 15 nodes, 10$\times$\texttt{nrf52dk} and 5$\times$\texttt{nrf52840dk}.
All nodes are within radio range of each other and are located in the same room in a 1~m $\times$ 1~m two dimensional grid.
This room is located on the ground floor of a typical office building without dedicated shielding, making it subject to radio interferences in the 2.4~GHz band used by, \eg Bluetooth, WiFi, DECT.

\paragraph{Software configuration} 
We use RIOT version \texttt{c739516} based on \textit{RIOT 2021.09} and NimBLE version \texttt{b9c20ad} based on \textit{NimBLE 1.4}.
We use the BLE default 1~MBit mode and, if not stated otherwise, the default parameters defined by RIOT and NimBLE.

For GNRC we enable \textit{6LoWPAN}~\cite{RFC-4944} as well as \texttt{gcoap} to support the \textit{Constrained Application Protocol}~(CoAP)~\cite{RFC-7252}.
We use the default GNRC packet buffer size of 6144~bytes and set the \texttt{gcoap} buffer size to 5120~bytes.
Router solicitations are turned off as they are not~needed.

The NimBLE configuration for \ipbleadv is based on the default configuration of \ipbleconn but with the major difference that extended advertisements are enabled.
Both are configured to provide an MTU of 1280~bytes while the NimBLE buffer size is configured to 8.9~Kbytes for both setups.
\autoref{sec:apx-platform} contains more details about the NimBLE configuration used in our experiments.

\paragraph{Scenarios}
We want to model network performance that resembles real-world deployments and also take the characteristics of low-power embedded hardware into account.
For this reason, we focus on analyzing network metrics on the application layer by generating network load by sending CoAP packets \cite{RFC-7252}.

In common IoT scenarios, numerous nodes periodically send their data to a gateway. 
To reflect this, we deploy multiple producers and a single consumer.
Producers periodically send data based on non-confirmable CoAP~PUT messages (\ie no application layer retransmission) to a consumer node.
To prevent burst traffic, producers add a random jitter to their periodic producer~interval.

Our experiments use a many-to-one scenario with a single consumer and 14 producer nodes.
The scenario is deployed in three different topologies: 1-hop star, 3-hop tree, and 14-hop line.
All topologies are created using static IP routes such that the consumer is located in the center (star), root (tree), or edge~(line).
More details on the topologies are presented in \autoref{sec:apx-iotlab}.

\begin{figure*}[t]
	\centering
	\includegraphics[width=1.0\linewidth]{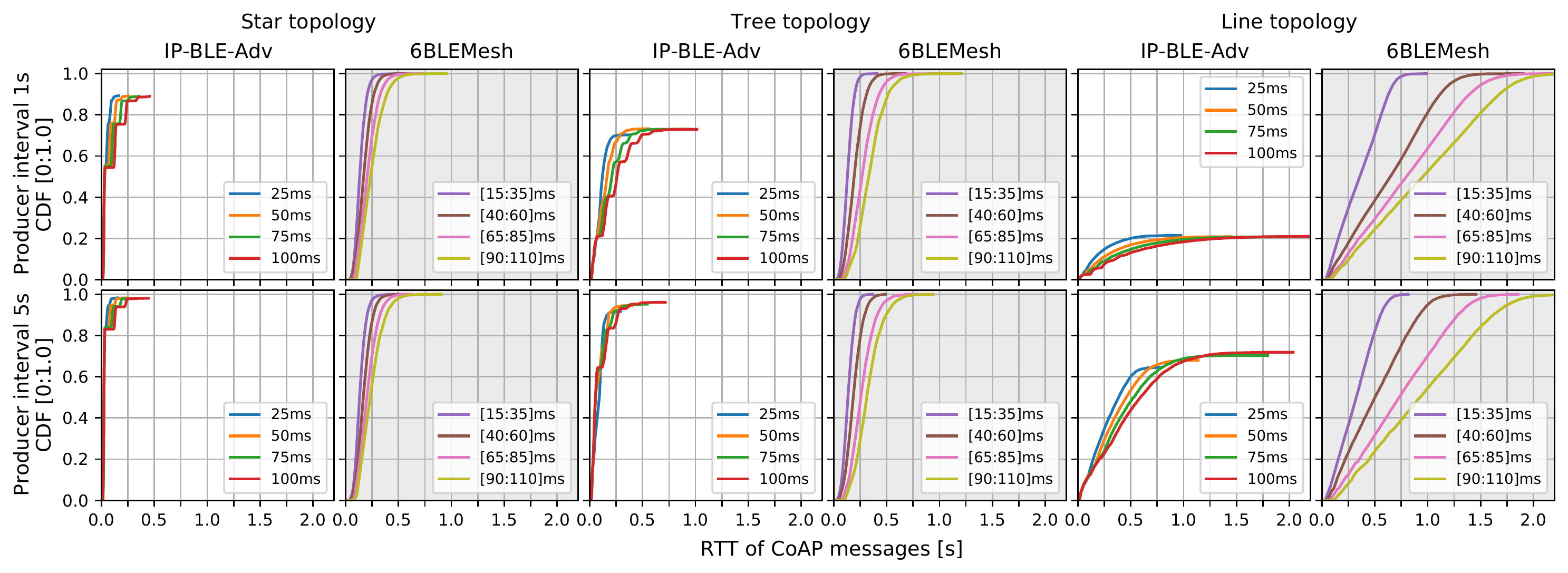}
	\caption{Distribution of CoAP round trip times for different producer intervals (1s, 5s) and varying advertising and connection intervals in three network topologies (star, tree, line).}
	\label{fig:perfoverview}
\end{figure*}

\section{Results}
\label{sec:results}

We compare network performance (\ie reliability and latency) and system performance (\ie power consumption and memory usage) for \ipbleadv and \ipbleconn in different network setups.

\subsection{Basic Performance Characteristics}
\label{sec:results_perf}

\autoref{fig:perfoverview} exhibits the latency of CoAP messages when deploying \ipbleadv and \ipbleconn in a star, tree, and line topology.
We measure the latency as the time difference between sending a CoAP message and receiving the corresponding (empty) ACK, which is sent even in non-confirmable mode.
Any packet loss leads to an infinite RTT.
We consider high and low network load (\ie producer interval of 1s~$\pm$.5s and 5s~$\pm$2.5s).
Each configuration runs for 1h.

In \ipbleadv experiments, we configure four different advertising intervals (25ms, 50ms, 75ms, 100ms) and a static packet retransmission of 2~packets, which results in 3 advertising events for each IP~packet (see \autoref{sec:results_params}).

In \ipbleconn experiments, to account for high reliability, we use randomized connection intervals \cite{psw-mgmio-21} such that these intervals match the advertising intervals used in \ipbleadv (\ie [15:35]ms, [40:60]ms, [65:85]ms, [90:100]ms).
To be independent of background noise, we conduct all \ipbleadv experiments outside of office hours.
We consider background noise in detail in \autoref{sec:results_resilience}.

\paragraph{\ipbleadv} 
The reliability of \ipbleadv differs greatly depending on the network topology and load but, in general, significantly decreases when the network load increases.
For example, using an advertising interval of 50ms and increasing the load from one packet every 5s to every 1s, reduces the CoAP packet delivery rates from 98.2\% to 89.1\%, 94.5\% to 73.2\%, and 67.9\% to 20.8\% in star, tree, and line topology, respectively.
The reason for these losses is twofold.
First, a higher network load increases the chance of packet collisions in the physical domain.
Second, more packets increase radio scheduling conflicts, especially at the consumer node.
In multi-hop tree and line topologies, these effects become multiplied because the overall number of packets sent is increased due to hop-wise packet forwarding along longer paths.
For example, in our setup, the overall number of packets sent among all nodes is $6\times$ higher in the line topology compared to the star topology.
Furthermore, even if we configured multi-hop topologies based on IP routes, packets are sent in the same radio domain.
Given that all nodes are in radio range of each other, packet collisions even via independent (IP) routes are observed.



The CoAP packet latency shows a stair effect.
These steps are caused by the static retransmissions of advertising packets and the width of the steps is defined by the advertising interval.
The delay transitions are most pronounced in the star topology.
In the tree and line topology, processing and queuing delays along intermediate nodes smoothen transitions.
In all configurations, we found that over 90\% of the successful CoAP acknowledgements are received after $2.5\times$ of the average hop count.

\paragraph{Comparison to \ipbleconn} 
The results of \ipbleconn (see \autoref{fig:perfoverview}) are in line with prior work~\cite{psw-mgmio-21}.
In all 24 experiments, no CoAP packet was lost.
The network load caused by a producer interval of 1s does not lead to an overload in \ipbleconn, only the delay slightly increases in multi-hop topologies.

In terms of reliability, \ipbleconn is superior compared to \ipbleadv.
Under relaxed network conditions, though, \ipbleadv achieves packet delivery rates that are acceptable for a wide range of IoT applications. 
In terms of latency, we observe a different picture: time-sliced channel hopping, which enables reliability in \ipbleconn, comes to the price of increased packet delay.
In \ipbleadv, packets are always sent immediately, \ie when they are handed to the Bluetooth Stack.
Then, packet delays are only affected by (relatively small) processing and retransmission delays.
Especially in scenarios with short paths (\eg star topology), \ipbleadv experiences significant shorter round-trip times compared to \ipbleconn.

\subsection{\ipbleadv Configuration Parameters}
\label{sec:results_params}

\begin{figure}[b]
	\centering
	\includegraphics[width=1.0\linewidth]{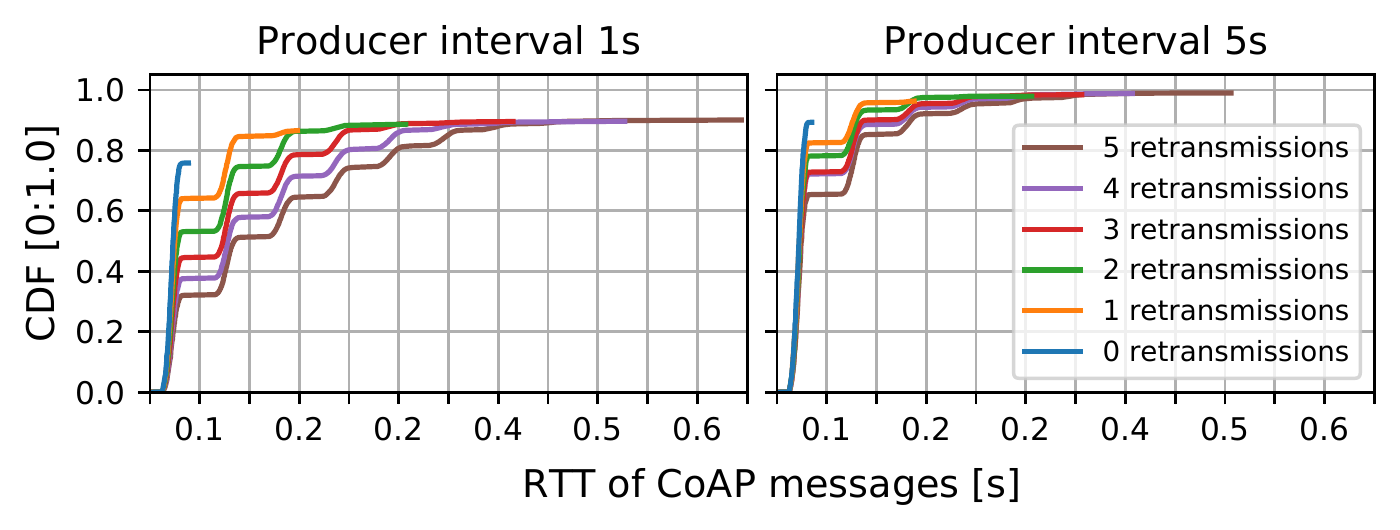}
	\caption{CoAP round trip time CDFs for different packet retransmission counts and producer intervals in a star topology network and an advertising interval of 50ms.}
	\label{fig:retransmission_impact}
\end{figure}

We identified two major configuration parameters that predominantly effect the network performance of \ipbleadv networks, the \textbf{retransmission count} and the \textbf{advertising interval}.

\paragraph{Retransmissions}
The unidirectional transmission of BLE advertisements challenges a reliable link layer because it prevents the implementation of acknowledgments to confirm successful messages. 
To increase the reliability of data transmissions, we apply a static retransmission scheme that defines how often an advertising event is replicated, similar to Bluetooth Mesh~\cite{bmp-bcsv-19}.

\autoref{fig:retransmission_impact} illustrates the impact of the retransmission, deployed in a star topology network and an advertising interval of 50ms.
A higher a number of retransmissions increases the overall reliability, as expected, but few retransmissions have surprisingly notable impact.
Focusing on the high load scenario (1s producer interval), the PDR significantly increases with a single retransmission (from 75.8\% to 86.5\%). 
Additional retransmissions have lower impact.
A PDR of 90.0\% requires 5~retransmissions.
This behavior can be reproduced in different topologies and under different network~loads.

The results in \autoref{fig:retransmission_impact} further illustrate a drawback of fixed retransmissions.
The reliability of the first advertising event reduces with the number of retransmissions and thus packet delays increase.
The reason for this is that nodes need to spend more time transmitting data while spending less time listening for incoming data, thus, the chance to miss incoming packets grows.
In our configuration, this can be observed particularly on the consumer node (not shown), which experiences the most IP traffic.
In our scenario, two retransmissions provide the best tradeoff between reliability and packet delay.

\paragraph{Advertising interval} The advertising interval defines the amount of time between two consecutive advertising events and thus the retransmission delay.
It is worth noting that a Bluetooth controller does not apply the configured value directly when scheduling advertising events, because the Bluetooth standard defines to add a random jitter of 0 to 10ms between two advertising events.
This jitter leads to less sharp transitions and a slight slope in the distribution of retransmissions (see \autoref{fig:retransmission_impact}).

\subsection{Noise Resilience}
\label{sec:results_resilience}
\begin{figure}[t]
	\begin{subfigure}[c]{.45\textwidth}
		\centering
		\includegraphics[width=1.0\linewidth]{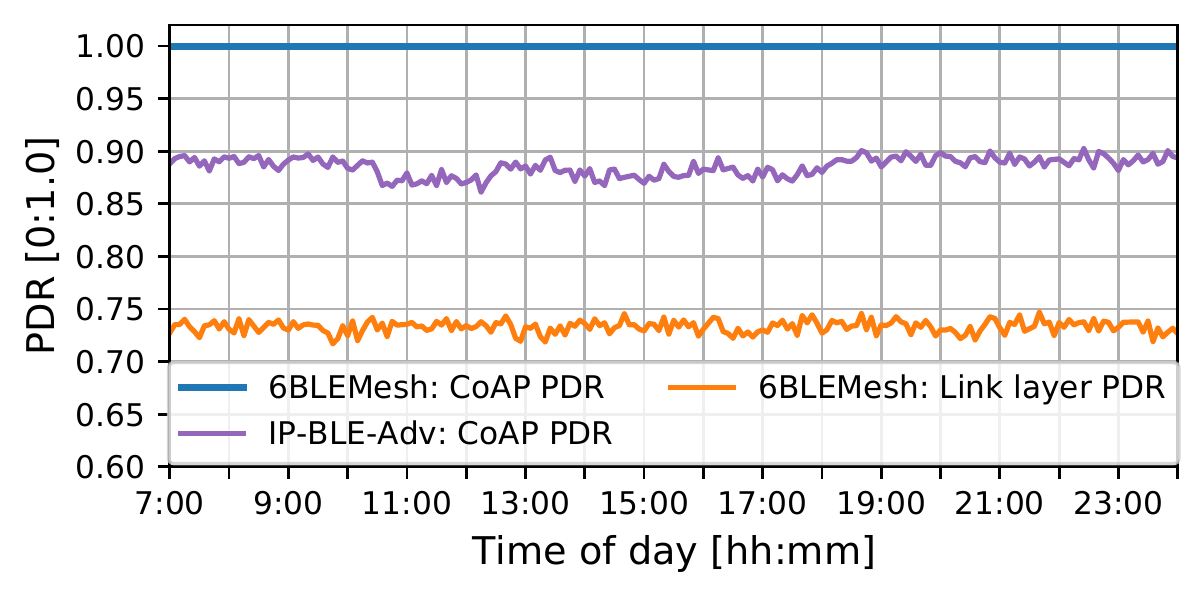}
		\caption{Packet delivery rates, binsize 300s.}
		\label{fig:24h_pdr}
	\end{subfigure}
	\begin{subfigure}[c]{.45\textwidth}
		 \includegraphics[width=1.0\linewidth]{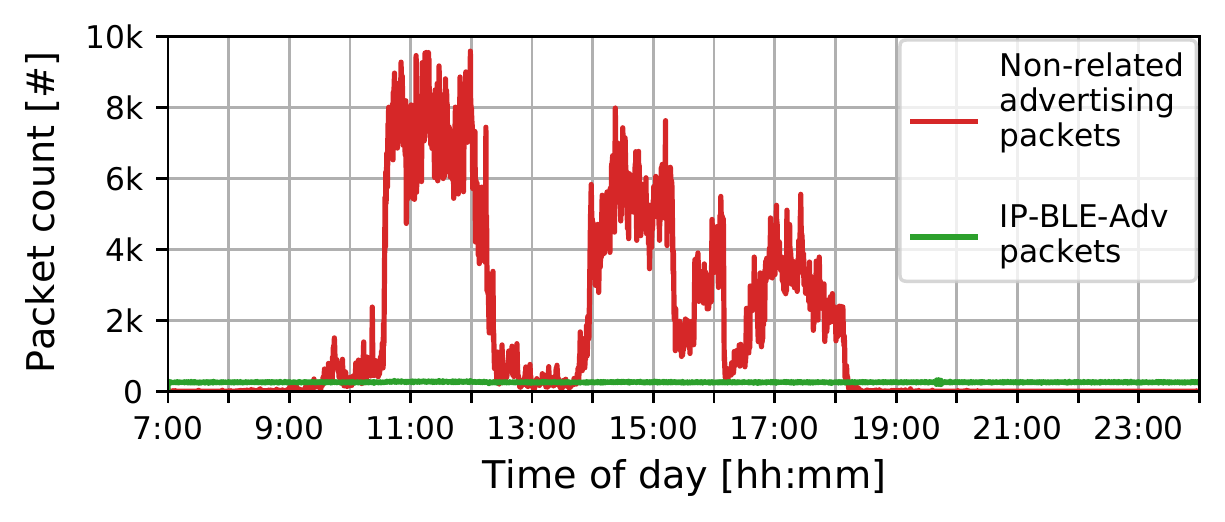}
		\caption{Total number of received advertising packets, binsize 10s.}
		\label{fig:24h_background}
	\end{subfigure}
	\caption{Impact of radio interference and background traffic on the network performance of \ipbleadv and \ipbleconn in star networks.}
\end{figure}

\ipbleadv uses the three primary advertising channels to deliver data.
These three channels can also be used by other BLE applications (\eg BLE beacons \cite{bbita-jssn-18}, Covid Warn Apps \cite{xccta-amxrm-20}) or interfere with external sources (\eg WiFi, DECT).
This leads to the assumption that \ipbleadv is less resilient to non-related radio activity compared to~\ipbleconn, which benefits from 37 typically less crowded data channels.

All used testbed nodes are located in an office building and are thus subject to diurnal radio background noise.
To illustrate the impact of such noise on the network performance, we conducted 18h~producer-consumer experiments in a star topology during working hours, 7AM -- 12AM (UTC+2).

\autoref{fig:24h_pdr} compares the CoAP packet delivery rates for both network setups and shows additionally the link layer PDR of the \ipbleconn network.
The \ipbleconn PDR is constant during the entire experiment and no impact of the external radio activities during office hours is visible.
The time-sliced channel hopping makes \ipbleadv robust against external interferences.

In \ipbleadv, the CoAP PDR exhibits a different picture: during office hours successful CoAP packet delivery decreases.
To shed light on the reason, \autoref{fig:24h_background} shows the aggregated number of unrelated BLE advertising packets received by all \ipbleadv nodes during the experiment.
Even though these packets are only a subset of the external noise, a correlation between increased noise and decreased CoAP PDR in the \ipbleadv network is visible.
We conclude that \ipbleadv is not critically sensitive to external noise (\ie 5\% additional packet loss) but environments with high radio activity have a negative impact on packet delivery rates.
This is not the case in \ipbleconn.

\subsection{Energy Consumption}
\label{sec:results_energy}
\begin{figure}[t]
	\begin{subfigure}[c]{.45\textwidth}
		\centering
		\includegraphics[width=1.0\linewidth]{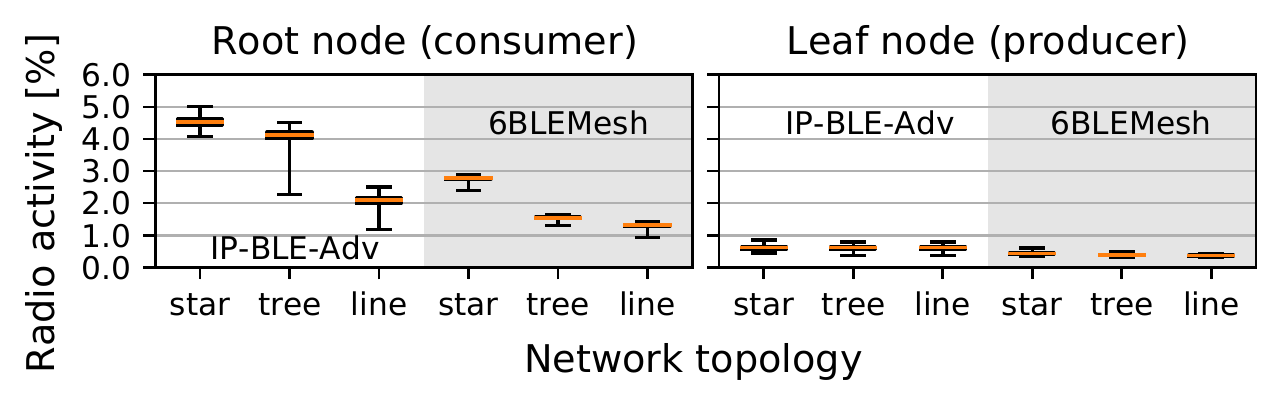}
		\caption{Radio usage for TX.}
		\label{fig:phyusage_tx}
	\end{subfigure}
	\begin{subfigure}[c]{.45\textwidth}
		 \includegraphics[width=1.0\linewidth]{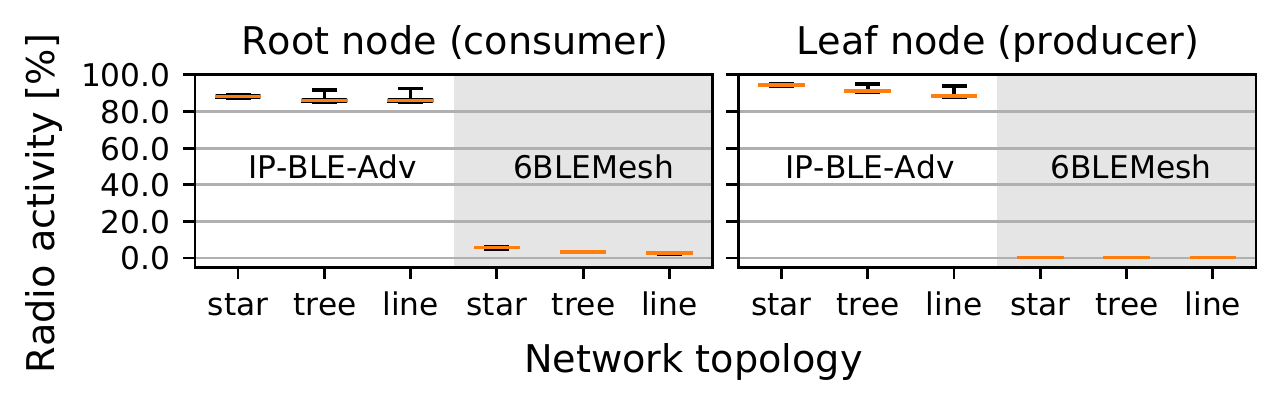}
		\caption{Radio usage for RX.}
		\label{fig:phyusage_rx}
	\end{subfigure}
	\caption{Comparison of radio utilization of \ipbleadv and \ipbleconn for the root node (consumer) and one selected leaf node (producer) and different network topologies. Advertising interval for \ipbleadv is 50ms, connection interval for \ipbleconn is [40:60]ms.}
	\label{fig:phyusage}
\end{figure}

Energy consumption is a key metric in low-power IoT networks.
As the FIT IoT-LAB does not provide any functions to measure the energy consumption directly, we derive this metric based on radio usage.
Assuming that a node is only active when sending or receiving network data, such as in our experiments, the radio activity does provide a close estimate of the actual energy consumption.
%
%
To measure the radio activity of each node, we count the time a radio is active in receive or transmit mode by inserting software counters directly in the low-level radio driver~code.

\autoref{fig:phyusage} shows the radio usage relative to the experiment runtime for the root of the network and a single selected leaf node.
All experiments have been running for 1h with an advertising interval of 50ms and 2 retransmissions in \ipbleadv and a connection interval of [40:60]ms in \ipbleconn.

\ipbleadv nodes experience more transmit events compared to the \ipbleconn~nodes (see \autoref{fig:phyusage_tx}).
Although in each network topology the same number of CoAP packets traverse the network, the overall number of BLE~packets is significantly larger in the \ipbleadv networks because multiple packets are needed for each advertising event and the static packet retransmission.
Differences become even more significant when measuring the amount of receiving time (see \autoref{fig:phyusage_rx}) since \ipbleadv requires that nodes always listen.
The aggregated radio activity of \ipbleadv nodes is below 100\%, though, due to radio switching and CPU overhead.

This means that \ipbleadv does not allow for low-power deployments.
If we assume the average current consumption of 4.6mA, given in the datasheet of our IoT hardware, the node lifetime would last at most 50h using a 230mAh coin cell battery.
In practice, this value would be smaller due to TX activity and energy consumed by CPU activity.
In comparison, the same node in a \ipbleconn network with a radio activity of 0.5\% has an average current consumption of 23µA and would last for 416 days on the same battery.

\begin{figure}[t]
  \begin{subfigure}{.45\textwidth}
    \centering
    \includegraphics[width=1.0\textwidth]{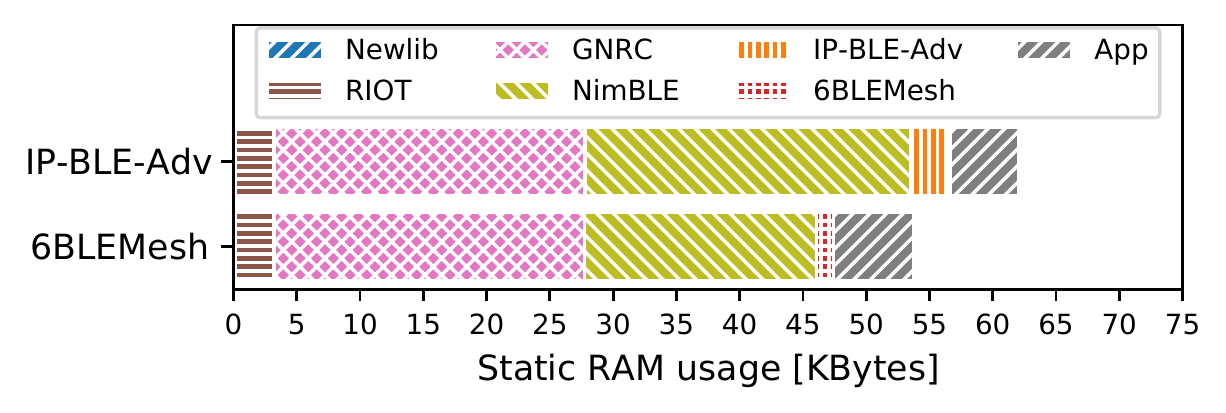}
    \caption{RAM requirements.}
    \label{fig:mem_ram}
  \end{subfigure}
  \begin{subfigure}{.45\textwidth}
    \centering
    \includegraphics[width=1.0\textwidth]{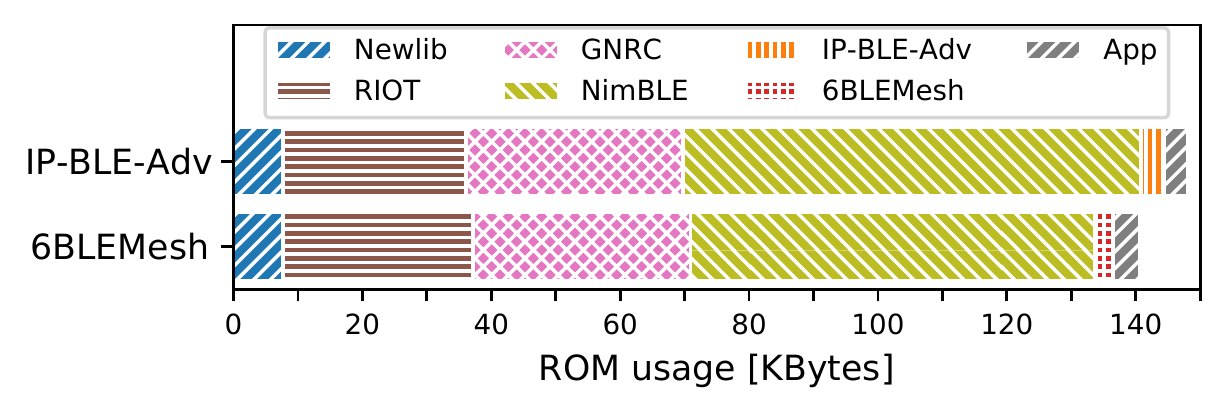}
    \caption{ROM requirements.}
    \label{fig:mem_rom}
  \end{subfigure}
  \caption{Compile-time memory usages of the \ipbleadv and \ipbleconn benchmark binaries, separated into different system components.}
  \label{fig:res_static_mem}
\end{figure}

\subsection{Memory Usage}
When analyzing memory requirements we consider two aspects to provide a complete picture: \one~the static RAM and ROM usage reserved and computed at linking time, and \two~the actual amount of RAM that is used at runtime.
Note that it is common to use only preallocated static memory in low-end embedded systems.
This is also the case in RIOT and NimBLE.

\autoref{fig:res_static_mem} compares the static memory usage between \ipbleadv and \ipbleconn.
Although the \ipbleadv Bluetooth stack is less complex, the resulting firmware image is 7.5~Kbyte larger compared to the \ipbleconn build.
The reason for this is that NimBLE currently cannot be configured to support extended advertisements without BLE connection handling functions.
Hence, enabling extended advertisements adds additional RAM and ROM usage to \ipbleadv, while \ipbleconn does not use that feature.

In \ipbleconn, the RAM usage depends on the preconfigured number of BLE connections a node can maintain simultaneously.
For each BLE connection an additional RAM block of 1056~bytes needs to be allocated.

We want to emphasize that the measured code sizes can be decreased significantly when moving to production code.
Our experiment applications contain a number of large software modules, such as the RIOT shell and our custom event logging, that can be dropped in user applications.
Furthermore, to limit the impact of overflowing buffers, the NimBLE and GNRC packet buffers are configured to fill the unused RAM space, which can also be decreased in practice.

\begin{figure}[t]
    \begin{subfigure}{.45\textwidth}
    \centering
    \includegraphics[width=1.0\textwidth]{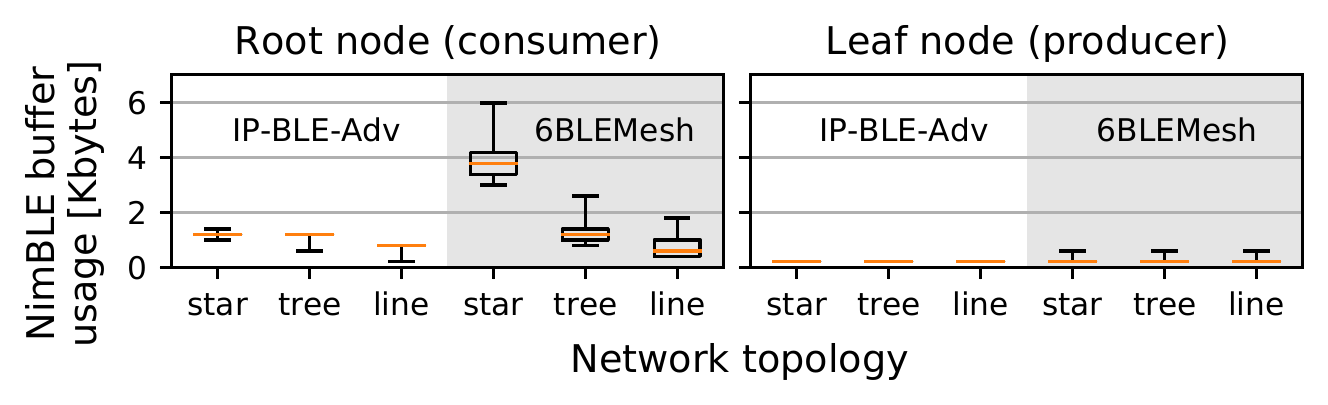}
    \caption{Producer interval 1s.}
    \label{fig:mem_1s}
  \end{subfigure}
  \begin{subfigure}{.45\textwidth}
    \centering
    \includegraphics[width=1.0\textwidth]{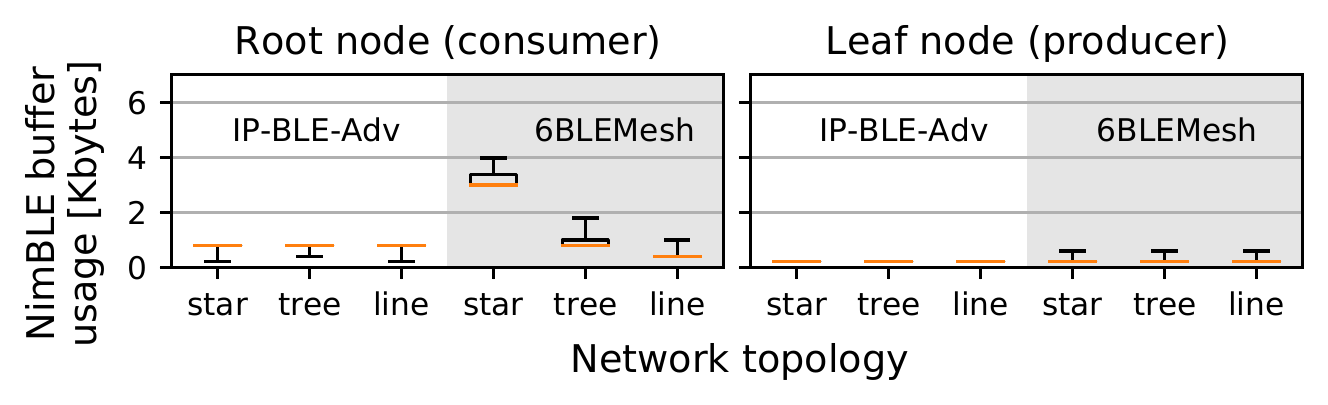}
    \caption{Producer interval 5s.}
    \label{fig:mem_5s}
  \end{subfigure}
  \caption{NimBLE packet buffer usages for a set of 1h experiments deployed in different network topologies, using an advertising interval of 50ms in \ipbleconn and 2 retransmissions and a connection interval of [40:60]ms in \ipbleadv.}
  \label{fig:res_mem_nimbuf}
\end{figure}

\autoref{fig:res_mem_nimbuf} exhibits the memory used in the NimBLE packet buffer during runtime of different 1h experiments.
We choose an advertising interval of 50ms with 2 retransmissions in the \ipbleadv setup and a connection interval of [40:60]ms in the \ipbleconn setup.
We show the buffer usage of the root node (consumer), which experiences the most network traffic, and a selected leaf node (one producer), experiencing the least traffic.
Other leaf nodes show the same behavior.

The buffer usage of the root node in the \ipbleadv network slightly increases with increasing network traffic.
In contrast to this, in \ipbleconn networks, the root node uses significant more buffer space, not only when the number of open connections grow (star vs. tree vs. line), but also when network load increases (producer interval 5s vs. 1s).
The buffer usages on the leaf node are comparable in both approaches but are slightly more volatile in \ipbleconn.

\section{Discussion}
\label{sec:discussion}

\paragraph{Bluetooth Mesh vs. \ipbleadv}
Similar to \ipbleadv, Bluetooth Mesh is connection-less but, in contrast, limited in exploiting BLE~channels and forwarding because of two restrictions.
First, Bluetooth Mesh uses only the primary advertising channels, whereas \ipbleadv leverages both three~primary and 37~secondary channels to reduce the load on the primary channels.
Second, Bluetooth Mesh floods packets, whereas \ipbleadv can leverage efficient IP-based routing.
Hence, \ipbleadv performance properties describe an upper bound for Bluetooth Mesh.
Furthermore, \ipbleadv allows for arbitrary IP~packets instead of small packets that follow a specific data model, granting more freedom when implementing Internet services and applications.

\paragraph{802.15.4 vs. \ipbleadv}
Our results show that \ipbleadv networks offer performance properties that are on par with common IEEE~802.15.4-based networks \cite{gklp-ncmcm-18} and thus sufficient to efficiently run advanced IoT protocols and applications in multi-hop environments.
Especially in scenarios of relaxed network loads and topologies with short paths, data transmission is reliable ($<$10\%~packet loss).
From a systems perspective, the biggest advantage of \ipbleadv compared to IEEE~802.15.4 is reduced complexity.
The same radio hardware and software can be used to implement both connection-less and connection-based communication.
Finally, \ipbleadv provides a higher data rate (8$\times$), which reduces transmission time and thus energy consumption.

\paragraph{\ipbleconn vs. \ipbleadv}
\ipbleconn outperforms \ipbleadv in terms of reliability (\autoref{sec:results_perf}) and energy consumption (\autoref{sec:results_energy}) but leads to higher packet latency.
The time sliced channel hopping in \ipbleconn allows for extremely efficient radio usage and reliable data transfer.
The management of allocating channels, however, adds radio and CPU overhead as well as latency.
We argue that \ipbleadv is a deployment option for IP over BLE applications that need to be optimized for low latency while being able to tolerate certain overhead in terms of energy consumption and packet loss.

When configuring \ipbleadv networks we identified the advertising interval and the retransmission count as major parameters influencing the network performance (\autoref{sec:results_params}).
The advertising interval is directly correlated to the latency of retransmitted packets while the retransmission count impacts reliability.
For the latter we found that a retransmission count of 2 events provides the best trade-off between increase in network load and reliability gains, where values above 2 did only marginally increase the reliability.
In the current state of the \ipbleadv design, where nodes use a scanning duty cycle of 100\% (radio in RX per default), the impacts of different parameters on the energy consumption are negligible.




\section{Related Work}
\label{sec:relwork}
Transferring data over BLE advertisements in multi-hop networks has been standardized by the Bluetooth SIG in Bluetooth Mesh~\cite{bmp-bcsv-19} and been extended by third-parties to support extended advertisements \cite{phgbv-sbm-21}.
Bluetooth Mesh uses proprietary network and data protocols and does not support transfer of IP data.
To the best of our knowledge, there is no prior work on transferring IP over BLE~advertisements.

Nikodem \etal~\cite{nb-eeabb-20} analyzed the reliability of BLE advertisements in a massive star topology deployment of 200 nodes.
They show that increasing advertising traffic does significantly reduce packet delivery rates, which is in line with our results.
In earlier work \cite{pkgsw-bmumh-19}, we showed that static packet retransmissions combined with flood-based routing applied by Bluetooth Mesh significantly multiplies the number of advertising packets sent.
This behavior does not apply in \ipbleadv networks due to IP-based routing instead of flooding.
The performance of CoAP in common large-scale IEEE~802.15.4 networks was analyzed by G{\"u}ndo{\u{g}}an~\etal \cite{gklp-ncmcm-18}.
\ipbleadv shows a similar network performance.

There is prior work on which potential optimizations of our initial design may rely on.
In our current \ipbleadv proposal, all nodes use a BLE scanner duty cycle of 100\% to keep the design and implementation lean and simple.
To improve energy consumption, we could adapt parts that focused on optimizing BLE advertising and scan parameters given measures such as reliability or energy~\cite{sedds-sh-21, bpcnd-jlmp-17, pmand-sr-20, gz-bbwne-21, bbepc-ypmn-20}.
Further optimization schemes to reduce radio interference \cite{hxcgl-acvbb-21, yt-sacps-20, gggt-eacwz-11} and network load \cite{nsb-ecsbl-20} in BLE advertising networks could potentially also be applied to optimize \ipbleadv networks.


\section{Conclusion}
\label{sec:conclusion}

In this work, we presented \ipbleadv, multi-hop IPv6 over BLE extended advertisements.
In contrast to other common connection-less, low-power IoT technologies such as IEEE~802.15.4 and Bluetooth Mesh, \ipbleadv features higher data rates and less energy consumption as well as higher throughput and the flexible implementation of Internet services and applications.
Our experiments based on real IoT hardware deployed in a mid-sized testbed show that \ipbleadv can complement IP over connection-based BLE~(\ipbleconn).
\ipbleadv networks offer low latency and reliability that is sufficient in many multi-hop deployments as long as the network load is relaxed.
\ipbleconn, on the other hand, is robust even in stressed networks and tailored to very low-power scenarios.
%
Therefore, \ipbleadv closes the gap towards a standard-compliant BLE~stack that provides both connection-less and connection-based communication in Internet scenarios.
We plan to analyze possible interferences due to concurrent operation and cross-dependencies in future research.

\label{lastpage}


\balance
\bibliographystyle{ACM-Reference-Format}
\bibliography{own,rfcs,ids,ngi,iot,mylit}


\begin{thebibliography}{39}


\ifx \showCODEN    \undefined \def \showCODEN     #1{\unskip}     \fi
\ifx \showDOI      \undefined \def \showDOI       #1{#1}\fi
\ifx \showISBNx    \undefined \def \showISBNx     #1{\unskip}     \fi
\ifx \showISBNxiii \undefined \def \showISBNxiii  #1{\unskip}     \fi
\ifx \showISSN     \undefined \def \showISSN      #1{\unskip}     \fi
\ifx \showLCCN     \undefined \def \showLCCN      #1{\unskip}     \fi
\ifx \shownote     \undefined \def \shownote      #1{#1}          \fi
\ifx \showarticletitle \undefined \def \showarticletitle #1{#1}   \fi
\ifx \showURL      \undefined \def \showURL       {\relax}        \fi
\providecommand\bibfield[2]{#2}
\providecommand\bibinfo[2]{#2}
\providecommand\natexlab[1]{#1}
\providecommand\showeprint[2][]{arXiv:#2}

\bibitem[\protect\citeauthoryear{??}{nim}{[n.d.]}]%
        {nimble-web}
 \bibinfo{year}{[n.d.]}\natexlab{}.
\newblock \bibinfo{booktitle}{\emph{{Apache NimBLE}}}.
\newblock
\urldef\tempurl%
\url{https://github.com/apache/mynewt-nimble}
\showURL{%
\tempurl}
\newblock
\shownote{\url{https://github.com/apache/mynewt-nimble}.}


\bibitem[\protect\citeauthoryear{??}{rio}{[n.d.]}]%
        {riot-web}
 \bibinfo{year}{[n.d.]}\natexlab{}.
\newblock \bibinfo{booktitle}{\emph{{RIOT OS}}}.
\newblock
\urldef\tempurl%
\url{https://www.riot-os.org/}
\showURL{%
\tempurl}
\newblock
\shownote{\url{https://www.riot-os.org/}.}


\bibitem[\protect\citeauthoryear{Adjih, Baccelli, Fleury, Harter, Mitton, Noel,
  Pissard-Gibollet, Saint-Marcel, Schreiner, Vandaele, and Watteyne}{Adjih
  et~al\mbox{.}}{2015}]%
        {abfhm-filso-15}
\bibfield{author}{\bibinfo{person}{Cedric Adjih}, \bibinfo{person}{Emmanuel
  Baccelli}, \bibinfo{person}{Eric Fleury}, \bibinfo{person}{Gaetan Harter},
  \bibinfo{person}{Nathalie Mitton}, \bibinfo{person}{Thomas Noel},
  \bibinfo{person}{Roger Pissard-Gibollet}, \bibinfo{person}{Frederic
  Saint-Marcel}, \bibinfo{person}{Guillaume Schreiner}, \bibinfo{person}{Julien
  Vandaele}, {and} \bibinfo{person}{Thomas Watteyne}.}
  \bibinfo{year}{2015}\natexlab{}.
\newblock \showarticletitle{{FIT IoT-LAB: A large scale open experimental IoT
  testbed}}. In \bibinfo{booktitle}{\emph{2015 IEEE 2nd World Forum on Internet
  of Things (WF-IoT)}}. \bibinfo{publisher}{IEEE Press},
  \bibinfo{address}{Piscataway, NJ, USA}, \bibinfo{pages}{459--464}.
\newblock


\bibitem[\protect\citeauthoryear{Ahmed, Michelin, Xue, Ruj, Malaney, Kanhere,
  Seneviratne, Hu, Janicke, and Jha}{Ahmed et~al\mbox{.}}{2020}]%
        {xccta-amxrm-20}
\bibfield{author}{\bibinfo{person}{Nadeem Ahmed}, \bibinfo{person}{Regio~A.
  Michelin}, \bibinfo{person}{Wanli Xue}, \bibinfo{person}{Sushmita Ruj},
  \bibinfo{person}{Robert Malaney}, \bibinfo{person}{Salil~S. Kanhere},
  \bibinfo{person}{Aruna Seneviratne}, \bibinfo{person}{Wen Hu},
  \bibinfo{person}{Helge Janicke}, {and} \bibinfo{person}{Sanjay~K. Jha}.}
  \bibinfo{year}{2020}\natexlab{}.
\newblock \showarticletitle{A Survey of COVID-19 Contact Tracing Apps}.
\newblock \bibinfo{journal}{\emph{IEEE Access}}  \bibinfo{volume}{8}
  (\bibinfo{year}{2020}), \bibinfo{pages}{134577--134601}.
\newblock
\urldef\tempurl%
\url{https://doi.org/10.1109/ACCESS.2020.3010226}
\showDOI{\tempurl}


\bibitem[\protect\citeauthoryear{Aly, Khomh, Guéhéneuc, Washizaki, and
  Yacout}{Aly et~al\mbox{.}}{2018}]%
        {akgwy-ftsit-18}
\bibfield{author}{\bibinfo{person}{M. Aly}, \bibinfo{person}{F. Khomh},
  \bibinfo{person}{Y. Guéhéneuc}, \bibinfo{person}{H. Washizaki}, {and}
  \bibinfo{person}{S. Yacout}.} \bibinfo{year}{2018}\natexlab{}.
\newblock \showarticletitle{{Is Fragmentation a Threat to the Success of the
  Internet of Things?}}
\newblock \bibinfo{journal}{\emph{IEEE Internet of Things Journal}}
  \bibinfo{volume}{14}, \bibinfo{number}{8} (\bibinfo{date}{Aug}
  \bibinfo{year}{2018}).
\newblock
\showISSN{2327-4662}
\urldef\tempurl%
\url{https://doi.org/10.1109/JIOT.2018.2863180}
\showDOI{\tempurl}


\bibitem[\protect\citeauthoryear{Baccelli, G{\"u}ndogan, Hahm, Kietzmann,
  Lenders, Petersen, Schleiser, Schmidt, and W{\"a}hlisch}{Baccelli
  et~al\mbox{.}}{2018}]%
        {bghkl-rosos-18}
\bibfield{author}{\bibinfo{person}{Emmanuel Baccelli}, \bibinfo{person}{Cenk
  G{\"u}ndogan}, \bibinfo{person}{Oliver Hahm}, \bibinfo{person}{Peter
  Kietzmann}, \bibinfo{person}{Martine Lenders}, \bibinfo{person}{Hauke
  Petersen}, \bibinfo{person}{Kaspar Schleiser}, \bibinfo{person}{Thomas~C.
  Schmidt}, {and} \bibinfo{person}{Matthias W{\"a}hlisch}.}
  \bibinfo{year}{2018}\natexlab{}.
\newblock \showarticletitle{{RIOT: an Open Source Operating System for Low-end
  Embedded Devices in the IoT}}.
\newblock \bibinfo{journal}{\emph{IEEE Internet of Things Journal}}
  \bibinfo{volume}{5}, \bibinfo{number}{6} (\bibinfo{date}{December}
  \bibinfo{year}{2018}), \bibinfo{pages}{4428--4440}.
\newblock
\urldef\tempurl%
\url{http://dx.doi.org/10.1109/JIOT.2018.2815038}
\showURL{%
\tempurl}


\bibitem[\protect\citeauthoryear{Bormann}{Bormann}{2014}]%
        {RFC-7400}
\bibfield{author}{\bibinfo{person}{C. Bormann}.}
  \bibinfo{year}{2014}\natexlab{}.
\newblock \bibinfo{booktitle}{\emph{{6LoWPAN-GHC: Generic Header Compression
  for IPv6 over Low-Power Wireless Personal Area Networks (6LoWPANs)}}}.
\newblock \bibinfo{type}{RFC} 7400. \bibinfo{institution}{IETF}.
\newblock


\bibitem[\protect\citeauthoryear{Bormann, Ersue, and Keranen}{Bormann
  et~al\mbox{.}}{2014}]%
        {RFC-7228}
\bibfield{author}{\bibinfo{person}{C. Bormann}, \bibinfo{person}{M. Ersue},
  {and} \bibinfo{person}{A. Keranen}.} \bibinfo{year}{2014}\natexlab{}.
\newblock \bibinfo{booktitle}{\emph{{Terminology for Constrained-Node
  Networks}}}.
\newblock \bibinfo{type}{RFC} 7228. \bibinfo{institution}{IETF}.
\newblock


\bibitem[\protect\citeauthoryear{Darroudi and Gomez}{Darroudi and
  Gomez}{2020}]%
        {ee6ib-dg-20}
\bibfield{author}{\bibinfo{person}{Seyed~Mahdi Darroudi} {and}
  \bibinfo{person}{Carles Gomez}.} \bibinfo{year}{2020}\natexlab{}.
\newblock \showarticletitle{{Experimental Evaluation of 6BLEMesh: IPv6-Based
  BLE Mesh Networks}}.
\newblock \bibinfo{journal}{\emph{Sensors}} \bibinfo{volume}{20},
  \bibinfo{number}{16} (\bibinfo{year}{2020}).
\newblock
\showISSN{1424-8220}
\urldef\tempurl%
\url{https://doi.org/10.3390/s20164623}
\showDOI{\tempurl}


\bibitem[\protect\citeauthoryear{Deering and Hinden}{Deering and
  Hinden}{2017}]%
        {RFC-8200}
\bibfield{author}{\bibinfo{person}{S. Deering} {and} \bibinfo{person}{R.
  Hinden}.} \bibinfo{year}{2017}\natexlab{}.
\newblock \bibinfo{booktitle}{\emph{{Internet Protocol, Version 6 (IPv6)
  Specification}}}.
\newblock \bibinfo{type}{RFC} 8200. \bibinfo{institution}{IETF}.
\newblock


\bibitem[\protect\citeauthoryear{Garroppo, Gazzarrini, Giordano, and
  Tavanti}{Garroppo et~al\mbox{.}}{[n.d.]}]%
        {gggt-eacwz-11}
\bibfield{author}{\bibinfo{person}{Rosario~G. Garroppo}, \bibinfo{person}{Loris
  Gazzarrini}, \bibinfo{person}{Stefano Giordano}, {and} \bibinfo{person}{Luca
  Tavanti}.} \bibinfo{year}{[n.d.]}\natexlab{}.
\newblock \showarticletitle{Experimental Assessment of the Coexistence of
  {{Wi}}-{{Fi}}, {{ZigBee}}, and {{Bluetooth}} Devices}. In
  \bibinfo{booktitle}{\emph{2011 {{IEEE International Symposium}} on a
  {{World}} of {{Wireless}}, {{Mobile}} and {{Multimedia Networks}}}}
  (2011-06). \bibinfo{pages}{1--9}.
\newblock
\urldef\tempurl%
\url{https://doi.org/10.1109/WoWMoM.2011.5986182}
\showDOI{\tempurl}


\bibitem[\protect\citeauthoryear{Geissdoerfer and Zimmerling}{Geissdoerfer and
  Zimmerling}{2021}]%
        {gz-bbwne-21}
\bibfield{author}{\bibinfo{person}{Kai Geissdoerfer} {and}
  \bibinfo{person}{Marco Zimmerling}.} \bibinfo{year}{2021}\natexlab{}.
\newblock \showarticletitle{Bootstrapping Battery-free Wireless Networks:
  Efficient Neighbor Discovery and Synchronization in the Face of
  Intermittency.}. In \bibinfo{booktitle}{\emph{NSDI}}.
  \bibinfo{pages}{439--455}.
\newblock


\bibitem[\protect\citeauthoryear{Gomez, Darroudi, Savolainen, and Spoerk}{Gomez
  et~al\mbox{.}}{2021a}]%
        {draft-ietf-6lo-blemesh}
\bibfield{author}{\bibinfo{person}{Carles Gomez}, \bibinfo{person}{Seyed
  Darroudi}, \bibinfo{person}{Teemu Savolainen}, {and} \bibinfo{person}{Michael
  Spoerk}.} \bibinfo{year}{2021}\natexlab{a}.
\newblock \bibinfo{booktitle}{\emph{{IPv6 Mesh over BLUETOOTH(R) Low Energy
  using IPSP}}}.
\newblock \bibinfo{type}{Internet-Draft -- work in progress}~10.
  \bibinfo{institution}{IETF}.
\newblock


\bibitem[\protect\citeauthoryear{Gomez, Darroudi, Savolainen, and Spörk}{Gomez
  et~al\mbox{.}}{2021b}]%
        {draft-gomez-6lo-blemesh-10}
\bibfield{author}{\bibinfo{person}{Carles Gomez}, \bibinfo{person}{Seyed
  Darroudi}, \bibinfo{person}{Teemu Savolainen}, {and} \bibinfo{person}{M.
  Spörk}.} \bibinfo{year}{2021}\natexlab{b}.
\newblock \bibinfo{booktitle}{\emph{{IPv6 Mesh over Bluetooth(R) Low Energy
  using IPSP}}}.
\newblock \bibinfo{type}{Internet-Draft -- work in progress}~10.
  \bibinfo{institution}{IETF}.
\newblock


\bibitem[\protect\citeauthoryear{Group}{Group}{2014}]%
        {bipsp-bcsv-14}
\bibfield{author}{\bibinfo{person}{Bluetooth Special~Interest Group}.}
  \bibinfo{year}{2014}\natexlab{}.
\newblock \bibinfo{booktitle}{\emph{{Internet Protocol Support Profile}}}.
\newblock \bibinfo{type}{Bluetooth Specification} 1.0.0.
  \bibinfo{institution}{Bluetooth SIG}.
\newblock
\urldef\tempurl%
\url{https://www.bluetooth.com/specifications/gatt/}
\showURL{%
\tempurl}


\bibitem[\protect\citeauthoryear{Group}{Group}{2016}]%
        {b50-bcsv-16}
\bibfield{author}{\bibinfo{person}{Bluetooth Special~Interest Group}.}
  \bibinfo{year}{2016}\natexlab{}.
\newblock \bibinfo{booktitle}{\emph{{Bluetooth Core Specification}}}.
\newblock \bibinfo{type}{Bluetooth Specification} 5.0.
  \bibinfo{institution}{Bluetooth SIG}.
\newblock
\urldef\tempurl%
\url{https://www.bluetooth.com/specifications/bluetooth-core-specification}
\showURL{%
\tempurl}


\bibitem[\protect\citeauthoryear{Group}{Group}{2019}]%
        {bmp-bcsv-19}
\bibfield{author}{\bibinfo{person}{Bluetooth Special~Interest Group}.}
  \bibinfo{year}{2019}\natexlab{}.
\newblock \bibinfo{booktitle}{\emph{{Bluetooth Mesh Profile}}}.
\newblock \bibinfo{type}{Mesh Profile} 1.0.1. \bibinfo{institution}{Bluetooth
  SIG}.
\newblock
\urldef\tempurl%
\url{https://www.bluetooth.com/specifications/specs/mesh-profile-1-0-1}
\showURL{%
\tempurl}


\bibitem[\protect\citeauthoryear{Group}{Group}{2020}]%
        {btmu-bcsv-20}
\bibfield{author}{\bibinfo{person}{Bluetooth Special~Interest Group}.}
  \bibinfo{year}{2020}\natexlab{}.
\newblock \bibinfo{booktitle}{\emph{{Bluetooth Market Update 2020}}}.
\newblock \bibinfo{type}{{T}echnical {R}eport}. \bibinfo{institution}{Bluetooth
  SIG}.
\newblock
\urldef\tempurl%
\url{https://www.bluetooth.com/bluetooth-resources/2020-bmu/}
\showURL{%
\tempurl}


\bibitem[\protect\citeauthoryear{Group}{Group}{2021}]%
        {bcss-bcsv-21}
\bibfield{author}{\bibinfo{person}{Bluetooth Special~Interest Group}.}
  \bibinfo{year}{2021}\natexlab{}.
\newblock \bibinfo{booktitle}{\emph{{Bluetooth Core Specification
  Supplement}}}.
\newblock \bibinfo{type}{Core Specification Supplement}~9.
  \bibinfo{institution}{Bluetooth SIG}.
\newblock
\urldef\tempurl%
\url{https://www.bluetooth.com/specifications/specs/core-specification-supplement-9}
\showURL{%
\tempurl}


\bibitem[\protect\citeauthoryear{G{\"u}ndogan, Kietzmann, Lenders, Petersen,
  Schmidt, and W{\"a}hlisch}{G{\"u}ndogan et~al\mbox{.}}{2018}]%
        {gklp-ncmcm-18}
\bibfield{author}{\bibinfo{person}{Cenk G{\"u}ndogan}, \bibinfo{person}{Peter
  Kietzmann}, \bibinfo{person}{Martine Lenders}, \bibinfo{person}{Hauke
  Petersen}, \bibinfo{person}{Thomas~C. Schmidt}, {and}
  \bibinfo{person}{Matthias W{\"a}hlisch}.} \bibinfo{year}{2018}\natexlab{}.
\newblock \showarticletitle{{NDN, CoAP, and MQTT: A Comparative Measurement
  Study in the IoT}}. In \bibinfo{booktitle}{\emph{Proc. of 5th ACM Conference
  on Information-Centric Networking (ICN)}}. \bibinfo{publisher}{ACM},
  \bibinfo{address}{New York, NY, USA}, \bibinfo{pages}{159--171}.
\newblock
\urldef\tempurl%
\url{https://doi.org/10.1145/3267955.3267967}
\showURL{%
\tempurl}


\bibitem[\protect\citeauthoryear{Han, Xu, Cao, Gao, and Lu}{Han
  et~al\mbox{.}}{[n.d.]}]%
        {hxcgl-acvbb-21}
\bibfield{author}{\bibinfo{person}{Daoqi Han}, \bibinfo{person}{Lingyi Xu},
  \bibinfo{person}{Ruohan Cao}, \bibinfo{person}{Hui Gao}, {and}
  \bibinfo{person}{Yueming Lu}.} \bibinfo{year}{[n.d.]}\natexlab{}.
\newblock \showarticletitle{Anti-{{Collision Voting Based}} on {{Bluetooth Low
  Energy Improvement}} for the {{Ultra}}-{{Dense Edge}}}.
\newblock   \bibinfo{volume}{9} (\bibinfo{year}{[n.\,d.]}),
  \bibinfo{pages}{73271--73285}.
\newblock
\showISSN{2169-3536}
\urldef\tempurl%
\url{https://doi.org/10.1109/ACCESS.2021.3079120}
\showDOI{\tempurl}


\bibitem[\protect\citeauthoryear{Hui and Thubert}{Hui and Thubert}{2011}]%
        {RFC-6282}
\bibfield{author}{\bibinfo{person}{J. Hui} {and} \bibinfo{person}{P. Thubert}.}
  \bibinfo{year}{2011}\natexlab{}.
\newblock \bibinfo{booktitle}{\emph{{Compression Format for IPv6 Datagrams over
  IEEE 802.15.4-Based Networks}}}.
\newblock \bibinfo{type}{RFC} 6282. \bibinfo{institution}{IETF}.
\newblock


\bibitem[\protect\citeauthoryear{Jeon, She, Soonsawad, and Ng}{Jeon
  et~al\mbox{.}}{2018}]%
        {bbita-jssn-18}
\bibfield{author}{\bibinfo{person}{Kang~Eun Jeon}, \bibinfo{person}{James She},
  \bibinfo{person}{Perm Soonsawad}, {and} \bibinfo{person}{Pai~Chet Ng}.}
  \bibinfo{year}{2018}\natexlab{}.
\newblock \showarticletitle{BLE Beacons for Internet of Things Applications:
  Survey, Challenges, and Opportunities}.
\newblock \bibinfo{journal}{\emph{IEEE Internet of Things Journal}}
  \bibinfo{volume}{5}, \bibinfo{number}{2} (\bibinfo{year}{2018}),
  \bibinfo{pages}{811--828}.
\newblock
\urldef\tempurl%
\url{https://doi.org/10.1109/JIOT.2017.2788449}
\showDOI{\tempurl}


\bibitem[\protect\citeauthoryear{{Julien}, {Liu}, {Murphy}, and
  {Picco}}{{Julien} et~al\mbox{.}}{2017}]%
        {bpcnd-jlmp-17}
\bibfield{author}{\bibinfo{person}{C. {Julien}}, \bibinfo{person}{C. {Liu}},
  \bibinfo{person}{A.~L. {Murphy}}, {and} \bibinfo{person}{G.~P. {Picco}}.}
  \bibinfo{year}{2017}\natexlab{}.
\newblock \showarticletitle{BLEnd: Practical Continuous Neighbor Discovery for
  Bluetooth Low Energy}. In \bibinfo{booktitle}{\emph{2017 16th ACM/IEEE
  International Conference on Information Processing in Sensor Networks
  (IPSN)}}. \bibinfo{pages}{105--116}.
\newblock


\bibitem[\protect\citeauthoryear{{Lee}, {Lee}, {Kim}, and {Bahk}}{{Lee}
  et~al\mbox{.}}{2016}]%
        {sarb-llkb-16}
\bibfield{author}{\bibinfo{person}{T. {Lee}}, \bibinfo{person}{M. {Lee}},
  \bibinfo{person}{H. {Kim}}, {and} \bibinfo{person}{S. {Bahk}}.}
  \bibinfo{year}{2016}\natexlab{}.
\newblock \showarticletitle{{A Synergistic Architecture for RPL over BLE}}. In
  \bibinfo{booktitle}{\emph{Proc. of 13th Annual IEEE International Conference
  on Sensing, Communication, and Networking (SECON)}}.
  \bibinfo{publisher}{IEEE}, \bibinfo{pages}{1--9}.
\newblock
\urldef\tempurl%
\url{https://doi.org/10.1109/SAHCN.2016.7732968}
\showDOI{\tempurl}


\bibitem[\protect\citeauthoryear{Lenders, Kietzmann, Hahm, Petersen,
  G{\"u}ndogan, Baccelli, Schleiser, Schmidt, and W{\"a}hlisch}{Lenders
  et~al\mbox{.}}{2018}]%
        {lkhpg-cwemr-18}
\bibfield{author}{\bibinfo{person}{Martine Lenders}, \bibinfo{person}{Peter
  Kietzmann}, \bibinfo{person}{Oliver Hahm}, \bibinfo{person}{Hauke Petersen},
  \bibinfo{person}{Cenk G{\"u}ndogan}, \bibinfo{person}{Emmanuel Baccelli},
  \bibinfo{person}{Kaspar Schleiser}, \bibinfo{person}{Thomas~C. Schmidt},
  {and} \bibinfo{person}{Matthias W{\"a}hlisch}.}
  \bibinfo{year}{2018}\natexlab{}.
\newblock \bibinfo{booktitle}{\emph{{Connecting the World of Embedded Mobiles:
  The RIOT Approach to Ubiquitous Networking for the Internet of Things}}}.
\newblock \bibinfo{type}{Technical Report} arXiv:1801.02833.
  \bibinfo{institution}{Open Archive: arXiv.org}.
\newblock
\urldef\tempurl%
\url{https://arxiv.org/abs/1801.02833}
\showURL{%
\tempurl}


\bibitem[\protect\citeauthoryear{Montenegro, Kushalnagar, Hui, and
  Culler}{Montenegro et~al\mbox{.}}{2007}]%
        {RFC-4944}
\bibfield{author}{\bibinfo{person}{G. Montenegro}, \bibinfo{person}{N.
  Kushalnagar}, \bibinfo{person}{J. Hui}, {and} \bibinfo{person}{D. Culler}.}
  \bibinfo{year}{2007}\natexlab{}.
\newblock \bibinfo{booktitle}{\emph{{Transmission of IPv6 Packets over IEEE
  802.15.4 Networks}}}.
\newblock \bibinfo{type}{RFC} 4944. \bibinfo{institution}{IETF}.
\newblock


\bibitem[\protect\citeauthoryear{Nieminen, Savolainen, Isomaki, Patil, Shelby,
  and Gomez}{Nieminen et~al\mbox{.}}{2015}]%
        {RFC-7668}
\bibfield{author}{\bibinfo{person}{J. Nieminen}, \bibinfo{person}{T.
  Savolainen}, \bibinfo{person}{M. Isomaki}, \bibinfo{person}{B. Patil},
  \bibinfo{person}{Z. Shelby}, {and} \bibinfo{person}{C. Gomez}.}
  \bibinfo{year}{2015}\natexlab{}.
\newblock \bibinfo{booktitle}{\emph{{IPv6 over BLUETOOTH(R) Low Energy}}}.
\newblock \bibinfo{type}{RFC} 7668. \bibinfo{institution}{IETF}.
\newblock


\bibitem[\protect\citeauthoryear{Nikodem and Bawiec}{Nikodem and
  Bawiec}{[n.d.]}]%
        {nb-eeabb-20}
\bibfield{author}{\bibinfo{person}{Maciej Nikodem} {and} \bibinfo{person}{Marek
  Bawiec}.} \bibinfo{year}{[n.d.]}\natexlab{}.
\newblock \showarticletitle{Experimental {{Evaluation}} of
  {{Advertisement}}-{{Based Bluetooth Low Energy Communication}}}.
\newblock  \bibinfo{volume}{20}, \bibinfo{number}{1}
  (\bibinfo{year}{[n.\,d.]}), \bibinfo{pages}{107}.
\newblock
Issue 1.
\urldef\tempurl%
\url{https://doi.org/10.3390/s20010107}
\showDOI{\tempurl}


\bibitem[\protect\citeauthoryear{Nikodem, Slabicki, and Bawiec}{Nikodem
  et~al\mbox{.}}{[n.d.]}]%
        {nsb-ecsbl-20}
\bibfield{author}{\bibinfo{person}{Maciej Nikodem}, \bibinfo{person}{Mariusz
  Slabicki}, {and} \bibinfo{person}{Marek Bawiec}.}
  \bibinfo{year}{[n.d.]}\natexlab{}.
\newblock \showarticletitle{Efficient {{Communication Scheme}} for {{Bluetooth
  Low Energy}} in {{Large Scale Applications}}}.
\newblock  \bibinfo{volume}{20}, \bibinfo{number}{21}
  (\bibinfo{year}{[n.\,d.]}), \bibinfo{pages}{6371}.
\newblock
Issue 21.
\urldef\tempurl%
\url{https://doi.org/10.3390/s20216371}
\showDOI{\tempurl}


\bibitem[\protect\citeauthoryear{Petersen, Kietzmann, G{\"u}ndogan, Schmidt,
  and W{\"a}hlisch}{Petersen et~al\mbox{.}}{2019}]%
        {pkgsw-bmumh-19}
\bibfield{author}{\bibinfo{person}{Hauke Petersen}, \bibinfo{person}{Peter
  Kietzmann}, \bibinfo{person}{Cenk G{\"u}ndogan}, \bibinfo{person}{Thomas~C.
  Schmidt}, {and} \bibinfo{person}{Matthias W{\"a}hlisch}.}
  \bibinfo{year}{2019}\natexlab{}.
\newblock \showarticletitle{{Bluetooth Mesh under the Microscope: How much ICN
  is Inside?}}. In \bibinfo{booktitle}{\emph{Proc. of 6th ACM Conference on
  Information-Centric Networking (ICN)}}. \bibinfo{publisher}{ACM},
  \bibinfo{address}{New York}, \bibinfo{pages}{134--140}.
\newblock
\urldef\tempurl%
\url{https://doi.org/10.1145/3357150.3357398}
\showURL{%
\tempurl}


\bibitem[\protect\citeauthoryear{Petersen, Schmidt, and W{\"a}hlisch}{Petersen
  et~al\mbox{.}}{2021}]%
        {psw-mgmio-21}
\bibfield{author}{\bibinfo{person}{Hauke Petersen}, \bibinfo{person}{Thomas~C.
  Schmidt}, {and} \bibinfo{person}{Matthias W{\"a}hlisch}.}
  \bibinfo{year}{2021}\natexlab{}.
\newblock \showarticletitle{{Mind the Gap: Multi-hop IPv6 over BLE in the
  IoT}}. In \bibinfo{booktitle}{\emph{Proc. of 17th International Conference on
  emerging Networking EXperiments and Technologies (CoNEXT))}}.
  \bibinfo{publisher}{ACM}, \bibinfo{address}{New York},
  \bibinfo{pages}{382--396}.
\newblock
\urldef\tempurl%
\url{https://doi.org/10.1145/3485983.3494847}
\showURL{%
\tempurl}


\bibitem[\protect\citeauthoryear{Pérez-Díaz-De-Cerio, Hernández-Solana,
  García-Lozano, Bardají, and Valenzuela}{Pérez-Díaz-De-Cerio
  et~al\mbox{.}}{[n.d.]}]%
        {phgbv-sbm-21}
\bibfield{author}{\bibinfo{person}{David Pérez-Díaz-De-Cerio},
  \bibinfo{person}{Ángela Hernández-Solana}, \bibinfo{person}{Mario
  García-Lozano}, \bibinfo{person}{Antonio~Valdovinos Bardají}, {and}
  \bibinfo{person}{José-Luis Valenzuela}.} \bibinfo{year}{[n.d.]}\natexlab{}.
\newblock \showarticletitle{Speeding {{Up Bluetooth Mesh}}}.
\newblock   \bibinfo{volume}{9} (\bibinfo{year}{[n.\,d.]}),
  \bibinfo{pages}{93267--93284}.
\newblock
\showISSN{2169-3536}
\urldef\tempurl%
\url{https://doi.org/10.1109/ACCESS.2021.3093102}
\showDOI{\tempurl}


\bibitem[\protect\citeauthoryear{Seo and Han}{Seo and Han}{2021}]%
        {sedds-sh-21}
\bibfield{author}{\bibinfo{person}{Jihun Seo} {and} \bibinfo{person}{Kijun
  Han}.} \bibinfo{year}{2021}\natexlab{}.
\newblock \showarticletitle{A Survey of Enhanced Device Discovery Schemes in
  Bluetooth Low Energy Networks}.
\newblock \bibinfo{journal}{\emph{IETE Technical Review}} \bibinfo{volume}{38},
  \bibinfo{number}{3} (\bibinfo{year}{2021}), \bibinfo{pages}{365--374}.
\newblock


\bibitem[\protect\citeauthoryear{Shan and Roh}{Shan and Roh}{2020}]%
        {pmand-sr-20}
\bibfield{author}{\bibinfo{person}{Gaoyang Shan} {and}
  \bibinfo{person}{Byeong-hee Roh}.} \bibinfo{year}{2020}\natexlab{}.
\newblock \showarticletitle{Performance Model for Advanced Neighbor Discovery
  Process in Bluetooth Low Energy 5.0-Enabled Internet of Things Networks}.
\newblock \bibinfo{journal}{\emph{IEEE Transactions on Industrial Electronics}}
  \bibinfo{volume}{67}, \bibinfo{number}{12} (\bibinfo{year}{2020}),
  \bibinfo{pages}{10965--10974}.
\newblock


\bibitem[\protect\citeauthoryear{Shelby, Hartke, and Bormann}{Shelby
  et~al\mbox{.}}{2014}]%
        {RFC-7252}
\bibfield{author}{\bibinfo{person}{Z. Shelby}, \bibinfo{person}{K. Hartke},
  {and} \bibinfo{person}{C. Bormann}.} \bibinfo{year}{2014}\natexlab{}.
\newblock \bibinfo{booktitle}{\emph{{The Constrained Application Protocol
  (CoAP)}}}.
\newblock \bibinfo{type}{RFC} 7252. \bibinfo{institution}{IETF}.
\newblock


\bibitem[\protect\citeauthoryear{Sp\"{o}rk, Boano, Zimmerling, and
  R\"{o}mer}{Sp\"{o}rk et~al\mbox{.}}{2017}]%
        {befpi-sbzr-17}
\bibfield{author}{\bibinfo{person}{Michael Sp\"{o}rk},
  \bibinfo{person}{Carlo~Alberto Boano}, \bibinfo{person}{Marco Zimmerling},
  {and} \bibinfo{person}{Kay R\"{o}mer}.} \bibinfo{year}{2017}\natexlab{}.
\newblock \showarticletitle{{BLEach: Exploiting the Full Potential of IPv6 over
  BLE in Constrained Embedded IoT Devices}}. In \bibinfo{booktitle}{\emph{Proc.
  of the 15th ACM Conference on Embedded Network Sensor Systems}} (Delft,
  Netherlands) \emph{(\bibinfo{series}{SenSys '17})}. \bibinfo{publisher}{ACM},
  \bibinfo{address}{New York, NY, USA}, Article \bibinfo{articleno}{2},
  \bibinfo{numpages}{14}~pages.
\newblock
\showISBNx{9781450354592}
\urldef\tempurl%
\url{https://doi.org/10.1145/3131672.3131687}
\showDOI{\tempurl}


\bibitem[\protect\citeauthoryear{Yang, Poellabauer, Mitra, and Neubecker}{Yang
  et~al\mbox{.}}{2020}]%
        {bbepc-ypmn-20}
\bibfield{author}{\bibinfo{person}{Jian Yang}, \bibinfo{person}{Christian
  Poellabauer}, \bibinfo{person}{Pramita Mitra}, {and} \bibinfo{person}{Cynthia
  Neubecker}.} \bibinfo{year}{2020}\natexlab{}.
\newblock \showarticletitle{Beyond beaconing: Emerging applications and
  challenges of BLE}.
\newblock \bibinfo{journal}{\emph{Ad Hoc Networks}}  \bibinfo{volume}{97}
  (\bibinfo{year}{2020}), \bibinfo{pages}{102015}.
\newblock
\showISSN{1570-8705}
\urldef\tempurl%
\url{https://doi.org/10.1016/j.adhoc.2019.102015}
\showDOI{\tempurl}


\bibitem[\protect\citeauthoryear{Yang and Tseng}{Yang and Tseng}{[n.d.]}]%
        {yt-sacps-20}
\bibfield{author}{\bibinfo{person}{Ting-Ting Yang} {and}
  \bibinfo{person}{Hsueh-Wen Tseng}.} \bibinfo{year}{[n.d.]}\natexlab{}.
\newblock \showarticletitle{A Service-Aware Channel Partition and Selection for
  Advertising in Bluetooth Low Energy Networks}. In
  \bibinfo{booktitle}{\emph{Proceedings of the 35th {{Annual ACM Symposium}} on
  {{Applied Computing}}}} ({New York, NY, USA}, 2020-03-30)
  \emph{(\bibinfo{series}{{{SAC}} '20})}. \bibinfo{publisher}{{Association for
  Computing Machinery}}, \bibinfo{pages}{2144--2150}.
\newblock
\showISBNx{978-1-4503-6866-7}
\urldef\tempurl%
\url{https://doi.org/10.1145/3341105.3373950}
\showDOI{\tempurl}


\end{thebibliography}

\appendix

\section{Artifacts}
\label{sec:apx-artifacts}
All artifacts used to create the results in this paper are publicly available.
These artifacts are composed of the actual implementation of \ipbleadv, tooling for conducting experiments in the FIT IoTlab testbed and analyzing their results, as well as all the raw output of all experiments.
Using these artifacts anyone should be able reproduce our results, not only
based on the provided raw data but also by re-running our experiments in an automated fashion.

Our \ipbleadv experiments are based on the implementation described in \autoref{sec:system_design}.
To support full reproducibility, the exact configuration parameters used in the conducted experiments are listed in \autoref{sec:eval} and \autoref{sec:apx-platform}.

\begin{figure*}[h]
    \begin{subfigure}[c]{.27\textwidth}
        \centering
        \includegraphics[width=.9\linewidth]{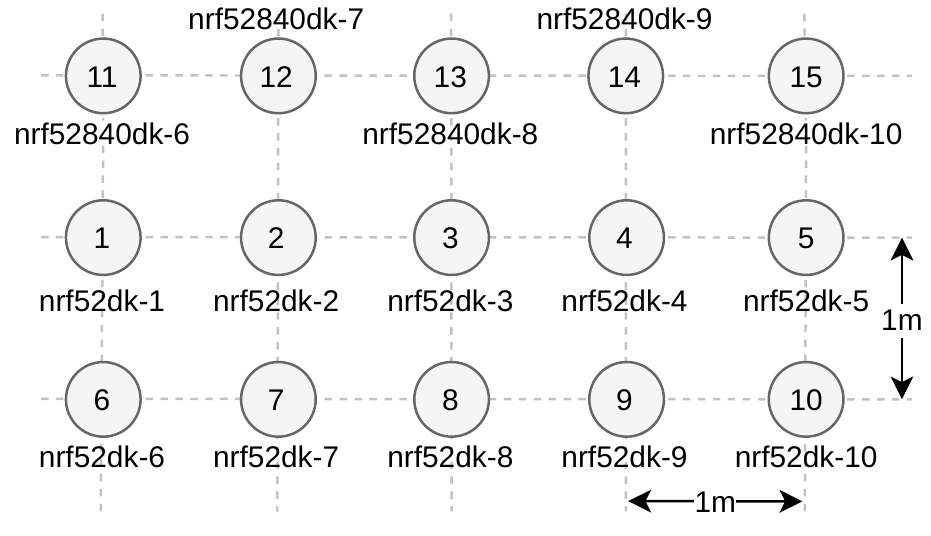}
        \caption{Spatial node distribution.}
        \label{fig:eval_topo_distribution}
    \end{subfigure}
    \begin{subfigure}[c]{.22\textwidth}
        \centering
        \includegraphics[width=.9\linewidth]{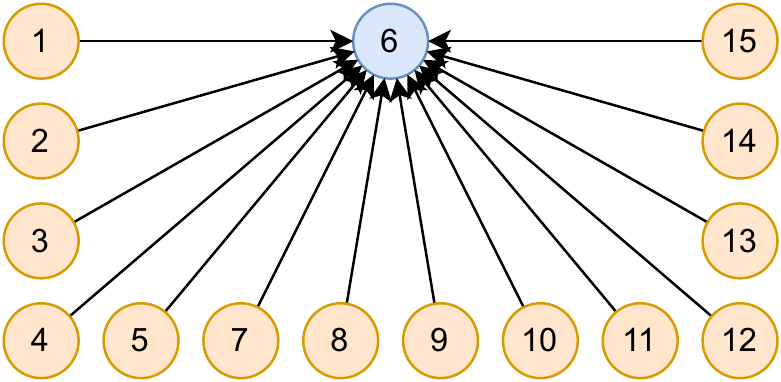}
        \caption{Star topology.}
        \label{fig:eval_topo_star}
    \end{subfigure}
    \begin{subfigure}[c]{.22\textwidth}
        \centering
        \includegraphics[width=.9\linewidth]{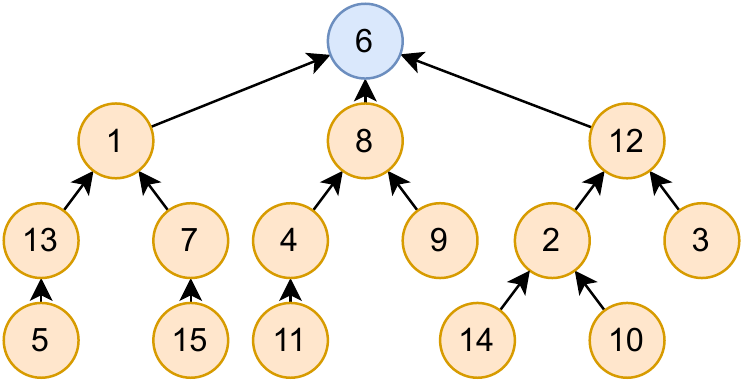}
        \caption{Tree topology.}
        \label{fig:eval_topo_tree}
    \end{subfigure}
    \begin{subfigure}[c]{.20\textwidth}
        \centering
        \includegraphics[width=.9\linewidth]{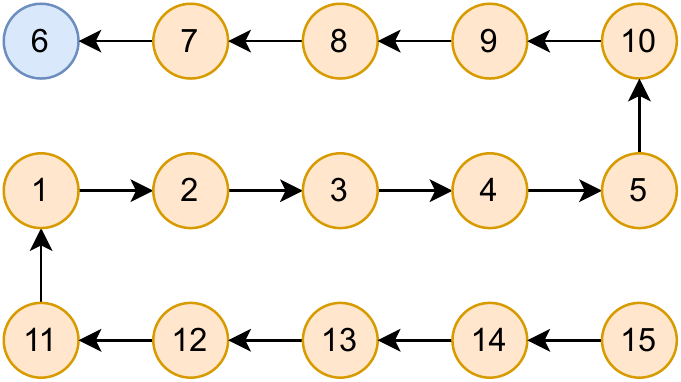}
        \caption{Line topology.}
        \label{fig:eval_topo_line}
    \end{subfigure}
    \caption{Mapping of nodes in the FIT IoTlab testbed to network topologies deployed in our experiments, highlighting producer~(orange) and consumer (blue) nodes.}
    \label{fig:eval_topo}
\end{figure*}

\subsection{Hosting}
\label{sec:apx-hosting}
All artifacts produced in this work are available through the following sources:
\begin{description}
\item[] \textbf{\url{https://github.com/ilabrg/artifacts-ccr-ipbleadv}}

Contains the detailed experiment descriptions and the tooling needed to run and analyze them.
As a starting point for reproducing our results, we recommend to read the instructions given in the \href{https://github.com/ilabrg/artifacts-ccr-ipbleadvv}{\texttt{README.md}}.

\item[] \textbf{\url{https://box.fu-berlin.de/s/2JJNWz5fDZ5jpyc}}

Contains the raw data used for creating the results presented in this work.
\end{description}

The final version of this paper will contain a link to a Zenodo (\url{https://zenodo.org}) archive for long-term accessibility.

\subsection{FIT IoTlab Setup}
\label{sec:apx-iotlab}
All raw data analyzed in this paper is the output of multi-node experiments conducted in the FIT IoTlab testbed (\url{https://www.iot-lab.info}).
The IoTlab is an open testbed and anyone can gain access by creating an account free of charge.

All experiments were run using 15 nRF52-based nodes at the Saclay site of the IoTlab.
We used 10 \textit{nrf52dk} nodes (\textit{nrf52dk-1} to \textit{nrf52dk-10}) and 5 \textit{nrf52840dk} nodes (\textit{nrf52840dk-6} to \textit{nrf52840dk-10}).
All nodes are located in the same room and evenly arranged in a 1m~$\times$~1m grid.
\autoref{fig:eval_topo} shows the spatial distribution of the nodes in the testbed as well as the mapping of these nodes to the deployed network topologies (star, tree, line).

\subsection{Software Platform}
\label{sec:apx-platform}
The software platform used for the implementation of \ipbleadv is built on RIOT \cite{riot-web} and NimBLE \cite{nimble-web}, which are both open source projects.
We contributed bug fixes to the underlying platforms in RIOT version \texttt{c739516} (based on \texttt{2021.07}) and NimBLE version \texttt{b9c20ad} (based on \textit{1.4}).

\ipbleadv is implemented as a module in RIOT, the implementation branch can be found at \url{https://github.com/haukepetersen/RIOT/tree/ipbleadv}.
The implementation of \ipbleadv is located in \texttt{pkg/nimble/jelling}.
The branch further contains a number hooks for collecting and printing trace data used for analyzing a network packet flow.

The NimBLE branch used in our experiments can be found at \url{https://github.com/haukepetersen/mynewt-nimble/tree/ipbleadv}.\\
This branch also contains a number of hooks to trace packets through the BLE stack.

Where applicable the default parameters provided in RIOT and NimBLE are used.
In addition to the basic configuration described in \autoref{sec:eval}, the following parameters are applied:

\paragraph{BLE}
In \ipbleconn, the BLE connections are statically setup to resemble the network topologies.
In both \ipbleadv and \ipbleconn, NimBLE is configured to allocate a packet buffer of 8.9~Kbytes.
Additionally, the data length extension is enabled in the controller.
The HCI interface is configured to transfer chunks up to 257~bytes, the maximum that NimBLE allows.

In the setup where we use \ipbleadv, extended advertisements are enabled and the maximum number of chained auxiliary packets per advertising event is configured to 10.
Furthermore, we allow a maximum number of 10 concurrent advertising instances.

\paragraph{Network}
To create the network topologies (see \autoref{fig:eval_topo}), IP routes are configured statically by using the RIOT shell commands.
In GNRC, the link layer packet queue is configured to hold 4~IP~packets.

\subsection{Experimentation Framework}
\label{sec:apx-expframework}
All experiments were controlled using a custom experimentation framework that takes care of allocating and controlling the nodes used in the testbed as well as collecting the experiment output.
In this framework, each experiment is fully described in a dedicated YAML configuration file.
Based on these configuration files it is possible to re-run any experiment.

The directory \href{https://github.com/ilabrg/artifacts-ccr-ipbleadv/tree/master/tools}{\texttt{tools/}} contains the tooling used to analyze the experiment results and to create the figures presented in this paper.

In the GitHub repository, the \href{https://github.com/ilabrg/artifacts-ccr-ipbleadv}{\texttt{README.md}}  contains more detailed descriptions of the framework and step-by-step instructions on its usage.


\end{document}